%% file: stellar_encounters_tgas.tex
\documentclass{aa}  
\usepackage{graphicx}
\usepackage{txfonts}
\usepackage{amsmath}
\usepackage{booktabs} % to typeset tables better
\usepackage{dcolumn}  % alignment on decimal point in tables
\usepackage{fixltx2e} % to keep table* in order with table
\usepackage[UKenglish]{isodate}
\cleanlookdateon % remove "th" in date format
\newcolumntype{d}[1]{D{.}{.}{#1}} % for dcolumn
\def\myeol{\\}

\def\tph{t_{\rm ph}}
\def\dph{d_{\rm ph}}
\def\vph{\mathrm v_{\rm ph}}

\def\tphlin{t_{\rm ph}^{\rm lin}}
\def\dphlin{d_{\rm ph}^{\rm lin}}

\def\tphmed{t_{\rm ph}^{\rm med}}
\def\dphmed{d_{\rm ph}^{\rm med}}
\def\vphmed{\mathrm v_{\rm ph}^{\rm med}}

\def\parallax{\varpi}

\def\propm{\mu}
\def\v{\mathrm v}
\def\vr{\mathrm v_r}
\def\sigvr{\sigma(\vr)}

\def\pmod{P_{\rm mod}}
\def\pexp{P^*_{\rm exp}}
\def\fmod{F_{\rm mod}}
\def\fexp{F_{\rm exp}}
\def\fobs{F_{\rm obs}}
\def\fint{F_{\rm int}}
\def\nmod{N_{\rm mod}}
\def\nexp{N_{\rm exp}}

\def\rlen{L}
\def\deriv{\mathrm d}
\def\given{\hspace{0.1em}|\hspace{0.1em}}

\def\deg{$^\circ$} 

\def\mas{mas}
\def\kms{km\,s$^{-1}$}
\def\ms{m\,s$^{-1}$}
\def\maspyr{\mas\,yr$^{-1}$}
\def\Msol{M$_\odot$}

\begin{document}

% See tgas_paper2/README for a summary of the procedure for producing the results in this paper.

%%%%%%%%%%%%%%%%%%%%%%%%%%%%%%%%%%%%%%%%%%%%%%%%%%%%%%%%%%%%

\title{The completeness-corrected rate of stellar encounters with the Sun from the first Gaia data release}
\titlerunning{Close encounters to the Sun in Gaia data release 1}
\author{C.A.L.~Bailer-Jones}
\institute{Max Planck Institute for Astronomy, K\"onigstuhl 17, 69117 Heidelberg, Germany. Email: calj@mpia.de}
\date{Submitted 27 June 2017; revised 4 August 2017; accepted 12 August 2017}
\abstract{ 
I report on close encounters of stars to the Sun found in the first Gaia data release (GDR1).  Combining Gaia astrometry with radial velocities of around 320\,000 stars drawn from various catalogues, I integrate orbits in a Galactic potential to identify those stars which pass within a few parsecs. Such encounters could influence the solar system, for example through gravitational perturbations of the Oort cloud.  16 stars are found to come within 2\,pc (although a few of these have dubious data). This is fewer than were found in a similar study based on Hipparcos data, even though the present study has many more candidates.
This is partly because I reject stars with large radial velocity uncertainties ($>$10\,\kms),
and partly because of missing stars in GDR1 (especially at the bright end). 
The closest encounter found is Gl\,710, a K dwarf long-known to come close to the Sun in about 1.3\,Myr. The Gaia astrometry predict a much closer passage than pre-Gaia estimates, however: just 16\,000\,AU (90\% confidence interval: 10\,000--21\,000\,AU), which will bring this star well within the Oort cloud. Using a simple model for the spatial, velocity, and luminosity distributions of stars, together with an approximation of the observational selection function, I model the incompleteness of this Gaia-based search as a function of the time and distance of closest approach.  Applying this to a subset of the observed encounters (excluding duplicates and stars with implausibly large velocities),
% i.e.\ with dph<5pc and |vr|<300 km/s.
I estimate the rate of stellar encounters within 5\,pc averaged over the past and future 5\,Myr to be $545 \pm 59$\,Myr$^{-1}$.  Assuming a quadratic scaling of the rate within some encounter distance (which my model predicts), this corresponds to $87 \pm 9$\,Myr$^{-1}$ within 2\,pc.
A more accurate analysis and assessment will be possible with future Gaia data releases.
}
\keywords{Oort cloud -- methods: analytical, statistical  -- solar neighbourhood -- stars: kinematics and dynamics -- surveys: Gaia} 
\maketitle

%%%%%%%%%%%%%%%%%%%%%%%%%%%%%%%%%%%%%%%%%%%%%%%%%%%%%%%%%%%%
\section{Introduction}

What influence has the Galactic environment had on the solar system? How has this environment changed with time?  As the solar system moves through the Galaxy, it is likely to come close to other stars. Ionizing radiation from particularly hot or active stars could adversely affect life on Earth. Tidal forces could shear the Oort cloud, the postulated body of primordial comets orbiting in the cold, dark outskirts of the solar system. This could push some bodies into the inner solar system in the form of comet showers, where they could potentially impact the Earth.

Reconstructing the encounter history (and future) of the solar system is possible by observing the present day positions and velocities of stars, and tracing their paths back and forward in time. Such studies came into their own after 1997 with the publication of the Hipparcos catalogue, which lists the positions, proper motions, and parallaxes of some 120\,000 stars. The various studies performed before and since Hipparcos have identified tens of stars which have -- or which will -- come within 2\,pc of the Sun \citep[][]{1994QJRAS..35....1M, 1996EM&P...72...19M, 1999AJ....117.1042G, 2001A&A...379..634G, 2006A&A...449.1233D, 2010AstL...36..220B, 2010AstL...36..816B, 2011MNRAS.418.1272J, 2015A&A...575A..35B, 2015MNRAS.449.2459D, 2015ApJ...800L..17M, 2016A&A...595L..10B}. One of these articles, \cite{2015A&A...575A..35B} (hereafter paper~1), is the precursor to the present work, and was based on the Hipparcos-2 catalogue \citep{2007ASSL..350.....V} combined with various radial velocity catalogues. \cite{2015MNRAS.449.2459D} looked more closely at some of the encounters discovered in paper~1, in particular those flagged as problematic.

The closer a stellar encounter, the larger the tidal force on the Oort cloud. But the ability to perturb the orbit of a comet depends also on the mass and relative speed of the encounter. This can be quantified using an impulse approximation
\citep{10.2307/20022899, 1950BAN....11...91O, 1976BAICz..27...92R, 1994CeMDA..58..139D}.
One version of this \citep{1976BAICz..27...92R} says the change in velocity of an Oort cloud comet due to a star of mass $M$ passing a distance $\dph$ at speed $\vph$ is proportional to
\begin{equation}
\frac{M}{\vph \, \dph^2} \ .
\label{eqn:impulse}
\end{equation}
For the much rarer, very close approaches (on the order of the comet--Sun separation),
% the distant case is Fig. 3 of Rickman 1976, section 4, point 2.
the dependence on distance is more like $1/\dph$ (see \cite{2015MNRAS.454.3267F} for some empirical comparison).
More sophisticated modelling of the effect of passing stars on the Oort cloud has been carried out by, for example,
\cite{1982A&A...112..157S, 1996EM&P...72...25W, 2002A&A...396..283D, 2011Icar..214..334F, 2012P&SS...73..124R}.
\cite{2015MNRAS.454.3267F} looked specifically at the consequence of the closest encounters found in paper~1.

Our ability to find and characterize close encounters has received a huge boost by the launch of the Gaia astrometric satellite in December 2013 \citep{2016A&A...595A...1G}. This deep, accurate astrometric survey will eventually provide positions, parallaxes, and proper motions for around a billion stars, and radial velocities for tens of millions of them. The first Gaia data release (GDR1) in September 2016 \citep{2016A&A...595A...2G}, although based on a short segment of data and with only preliminary calibrations, included astrometry for two million Tycho sources \citep{2000A&A...355L..27H} down to about 13th magnitude, with parallaxes and proper motions a few times more accurate than Hipparcos \citep{2016A&A...595A...4L}. This part of GDR1 is known as the Tycho-Gaia Astrometric Solution (TGAS), as it used the Tycho coordinates from epoch J1991.25 to lift the degeneracy between parallax and proper motion which otherwise would have occurred when using a short segment of Gaia data.

The first goal of this paper is to find encounters from combining TGAS with various radial velocity catalogues. The method, summarized in section~\ref{sec:procedure} is very similar to that described in paper~1.  One important difference here is the exclusion of stars with large radial velocity uncertainties.  I search for encounters out to 10\,pc, but really we are only interested in encounters which come within about 2\,pc. More distant encounters would have to be very massive and slow to significantly perturb the Oort cloud.  The encounters are presented and discussed in section~\ref{sec:results}. 

The other goal of this paper is to model the completeness in the perihelion time and distance of searches for close encounters. This involves modelling the spatial, kinematic, and luminosity distribution of stars in the Galaxy, and combining these with the survey selection function to construct a two-dimensional completeness function. A simple model is introduced and its consequences explored in section~\ref{sec:completeness}. Using this model I convert the number of encounters detected into a prediction of the intrinsic encounter rate out to some perihelion distance. This study is a precursor to developing a more sophisticated incompleteness-correction for subsequent Gaia data releases, for which the selection function should be better defined.

The symbols $\tph$, $\dph$, and $\vph$ indicate the perihelion time, distance, and speed respectively (``ph'' stands for perihelion). The superscript ``lin'' added to these refers to the quantity as found by the linear motion approximation of the nominal data (where ``nominal data'' means the catalogue values, rather than resamplings thereof). The superscript ``med'' indicates the median of the distribution over a perihelion parameter found by orbit integration (explained later).  Preliminary results of this work were reported at the Nice IAU symposium in April 2017 \citep{2017IAUS..331..xxxB}.

%%%%%%%%%%%%%%%%%%%%%%%%%%%%%%%%%%%%%%%%%%%%%%%%%%%%%%%%%%%%
\section{Procedure for finding close encounters}\label{sec:procedure}

The method is similar to that described in paper~1, but differs in part because the radial velocity (RV) catalogue was prepared in advance of the Gaia data release.

\begin{enumerate}

\item I searched CDS/Vizier in early 2016 for RV catalogues which had a magnitude overlap with Tycho-2 and a typical radial velocity precision, $\sigvr$, better than a few \kms. I ignored catalogues of stellar cluster members, those obsoleted by later catalogues, or those with fewer than 200 entries. Where necessary, I cross-matched them with Tycho-2 using the CDS X-match service to assign Tycho-2 IDs
(or Hipparcos IDs for those few Hipparcos stars which do not have Tycho-2 IDs). 
% The comment in parentheses actually refers to the separate step to match the RV catalogues via Hipparcos number or position to my list of 403 "Hipparcos stars with no Tycho-2 ID". This is explained in kinematic_catalogue/missing_hip/README.
The union of all these catalogues I call {\em RVcat}. A given star may appear in more than one RV catalogue or more than once in a given catalogue, so RVcat contains duplicate stars. Each unique entry in RVcat I refer to as an {\em object}.
RVcat contains 412\,742 objects.
% unchanged by the addition of the missing Hipparcos IDs

Missing $\sigvr$ values were replaced with the median of all the other values of $\sigvr$ in that catalogue.  Malaroda2012 lists no uncertainties, so I somewhat arbitrarily set them to 0.5\,\kms.  For APOGEE2 I took the larger of its catalogue entries {\tt RVscatter} and {\tt RVerrMed} and added 0.2\,\kms\ in quadrature, following the advice of Friedrich Anders (private communication). For Galah I conservatively use 0.6\,\kms, following the statement in \cite{2017MNRAS.465.3203M} that 98\% of all stars have a smaller standard deviation than this.

\item I then independently queried the Gaia-TGAS archive to find all stars which, assuming them all to have radial velocities of $|\vr|=750$\,\kms, would come within 10\,pc of the Sun according to the linear motion approximation (LMA) ($\dphlin<10$\,pc).
The LMA models stars as moving on unaccelerated paths relative to the Sun, so has a quick analytic solution (see paper~1).
By using a fixed radial velocity I could run the 
archive query (listed in appendix \ref{appendix:query}) independently of any RV catalogue, and by using the largest plausible radial velocity I obtained the most-inclusive list of encounters 
(the larger the radial velocity, the closer the approach).
% Using the actual radial velocity later on (see below) gives a larger perihelion distance, so can only remove stars. 
The value of 750\,\kms\ is chosen on the basis that the escape velocity from the Galaxy (with the model adopted for the orbit integrations) is 621\,\kms\ at the Galactic Centre (GC) and 406\,\kms\ at the solar circle; 
a reasonable upper limit is 500\,\kms. To this I add the velocity of the Sun relative to the GC (241\,\kms) to set a limit of 750\,\kms, on the assumption that very few stars are really escaping from the Galaxy.
This query (run on 16-11-2016) identified 541\,189 stars. 
Stars with non-positive parallaxes are removed,
% done in make_kinematic_catalogue.R
leaving 540\,883 stars (call this {\em ASTcat}). 
% not 484715 (see make_kinematic_catalogue.R).
% The above was 540480 when the 403 Hipparcos stars with no TychoID were not included (more precisely, no TychoID listed in my join of TGAS with Tycho-2).

\item I now remove from RVcat all objects which have $|\vr|>750$\,\kms\ or $\sigvr>10$\,\kms\ or which have one of these values missing. This reduces RVcat to 397\,788 objects, which corresponds to 322\,462 unique stars. 
% done in make_kinematic_catalogue.R. 
This is approximately the number of objects/stars I can test for being encounters.\footnote{\label{foot:number_of_candidates}This number is a slight overestimate because although all of these stars have Tycho/Hipparcos IDs and valid radial velocities, (1) not all are automatically in TGAS (due to the its bright magnitude limit and other omissions), and (2) objects with negative parallaxes are excluded.}
The number of objects for each of the catalogues is shown in column 3 of Table \ref{tab:RVcatalogues}.

\begin{table}
\begin{center}
\caption{The number of objects (not unique stars) in the RV catalogues.
The first column gives the catalogue reference number (used in Table \ref{tab:periStats}) and the second column a reference name. The third column lists the number of objects with valid Tycho/Hipparcos IDs which also have $|\vr|\leq750$\,\kms\ and $\sigvr\leq10$\,\kms. 
This is approximately the number of objects available for searching for encounters.
The fourth column lists how many of these have $\dphlin < 10$\,pc.
% Third and fourth columns are the number of objects in tgas_sel02_rvcat20161116.Robj, as listed at the end of make_kinematic_catalogue.R,
% *plus* the missing Hipparcos stars for each listed in kinematic_catalogue/missing_hip/README.
There are duplicate stars between (and even within) the catalogues.
\label{tab:RVcatalogues}}
\begin{tabular}{ r r r r }
\toprule
cat & name & \#Tycho & \#($\dphlin<10$\,pc)\\
\midrule
  1 & RAVE-DR5 & 302\,371  & 240  \\
  2 & GCS2011 &  13\,548   & 157  \\
  3 & Pulkovo & 35\,745  &  238  \\
  4  & Famaey2005 &  6\,047  & 1 \\
  5  & Web1995 &   429   & 5  \\
  6  & BB2000 &   670   &  5  \\
  7 &  Malaroda2012 &   1\,987  & 12  \\
  8  & Maldonado2010  &  349  & 32  \\
  9  &  Duflot1997 &   43    &  0 \\
 10 &  APOGEE2 &  25\,760 & 33  \\
 11  & Gaia-ESO-DR2   &  159   & 0  \\
 12 & Galah & 10\,680  &   2  \\
\midrule
Total & & 397\,788 & 725 \\
\bottomrule
\end{tabular}\\
{\bf Notes:} References and (where used) CDS catalogue numbers for the RV catalogues: (1) \cite{2017AJ....153...75K}; (2) \cite{2011A&A...530A.138C} {\tt J/A+A/530/A138/catalog}; (3) \cite{2006AstL...32..759G} {\tt III/252/table8}; (4) \cite{2005A&A...430..165F} {\tt J/A+A/430/165/tablea1}; (5) \cite{1995A&AS..114..269D} {\tt III/190B}; (6) \cite{2000A&AS..142..217B} {\tt III/213}; (7) \cite{2000A&AS..144....1M} {\tt III/249/catalog}; (8) \cite{2010A&A...521A..12M} {\tt J/A+A/521/A12/table1}; (9) \cite{1997A&AS..124..255F} {\tt J/A+AS/124/255/table1}; (10) \cite{2016arXiv160802013S}; (11) {\tt https://www.gaia-eso.eu}; (12) \cite{2017MNRAS.465.3203M}.
\end{center}
\end{table}

\item The common objects between this reduced RVcat and ASTcat are identified (using the Tycho/Hipparcos IDs). These 98\,849 
% was 98714 in submitted version. To this we add the 135 missing hip ids. 
objects all have -- by construction -- a Tycho/Hipparcos ID, TGAS data, positive parallaxes, $|\vr| \leq 750$\,\kms, $\sigvr \leq 10$\,\kms, and {\em would} encounter the Sun within 10\,pc (using the LMA) if they had a radial velocity of 750\,\kms.
Applying the LMA now with the measured radial velocity, I find that 725 have $\dphlin<10$\,pc. The number per RV catalogue is shown in the fourth column of Table \ref{tab:RVcatalogues}.
Only these objects will be considered in the rest of this paper.
The distribution of the measurements, including their standard deviations and correlations, can be found in \cite{2017IAUS..331..xxxB}.

\item \label{item:surrogates} These 725 objects are then subject to the orbit integration procedure described in sections 3.3 and 3.4 of paper~1. 
This gives accurate perihelia both by modelling the acceleration of the orbit and by numerically propagating the uncertainties through the nonlinear transformation of the astrometry.
To summarize: the 6D Gaussian distribution of the data (astrometry plus radial velocity) for each object is resampled to produce 2000  ``surrogate'' measurements of the object. These surrogates reflect the uncertainties in the position and velocity (and takes into account their covariances, which can be very large). The orbit for every surrogate is integrated through the Galactic potential and the perihelia found. The resulting distribution of the surrogates over the perihelion parameters represents the propagated uncertainty. 
These distributions are asymmetric; ignoring this can be fatal (see section \ref{sec:otherstudy} for an example).
I summarize the distributions using the median and the 5\% and 95\% percentiles (which together form an asymmetric 90\% confidence interval, CI).\footnote{In paper~1 I reported the mean, although the median is available in the 
online supplement: \url{http://www.mpia.de/homes/calj/stellar_encounters/stellar_encounters_1.html}. The median is more logical when reporting the equal-tailed confidence interval, as this interval is guaranteed to contain the median. The mean and median are generally very similar for these distributions.} I use the same Galactic model as in paper~1. The integration scheme is also the same, except that I now recompute the sampling times for each surrogate, as this gives a better sampling of the orbit.
%and so avoids errors which can occur when the measured radial velocity has low significance (meaning that some of the surrogates can move in the opposite direction).

\end{enumerate}

This procedure does not find all encounters, because an initial LMA-selection is not guaranteed to include all objects which, when properly integrated, would come within 10\,pc. Yet as the LMA turns out to be a reasonable approximation for most stars (see next section), this approach is adequate for identifying encounters which come much closer than 10\,pc. 

%%%%%%%%%%%%%%%%%%%%%%%%%%%%%%%%%%%%%%%%%%%%%%%%%%%%%%%%%%%%
\section{Close encounters found in TGAS}\label{sec:results}

% Plots in this section are made by stellar_encounters/tgas_paper2/analyse_results.R,
% specifically the block of code at the end of that file.

As duplicates are present in the RV catalogue, the 725 objects found correspond to 468 unique stars (unique Tycho/Hipparcos IDs). The numbers coming within various perihelion distances are shown in Table~\ref{tab:dphSummary}.  The data for those with $\dphmed < 2$\,pc are shown in Table~\ref{tab:periStats} (the online table at CDS includes all objects with $\dphmed < 10$\,pc).  Negative times indicate past encounters.

\begin{table}
\begin{center}
\caption{The number of objects and stars found by the orbit integration to have $\dphmed < \dph^{\rm max}$.
An object is a specific catalogue entry for a star. Objects with potentially problematic data have not been excluded.
\label{tab:dphSummary}}
\begin{tabular}{ d{1} r r }
\toprule
\multicolumn{1}{c}{$\dph^{\rm max}$}  & No.\ objects & No.\ stars \\
\midrule
\infty & 725 & 468 \\
10 & 646 & 402 \\
5 & 149 & 97 \\
3 & 56 & 42 \\
2 & 20 & 16 \\
1 & 2 & 2 \\
0.5 & 1 & 1 \\
\bottomrule
\end{tabular}
\end{center}
\end{table}

\begin{table*}
\centering
\tiny{
\caption{Perihelion parameters for all objects with a median perihelion distance (median of the samples; $\dphmed$) below 2\,pc, sorted by this value.
The first column list the Tycho ID. The second column indicates the RV catalogue (see Table \ref{tab:RVcatalogues}). Columns 3, 6, and 9 are
$\tphmed$, $\dphmed$, and $\vphmed$ respectively. The columns labelled 5\% and 95\% are the bounds of corresponding confidence intervals.
Columns 12--14 (``\% samples'') indicate the percentage of surrogates for each object for which $\dph$ is below 0.5, 1.0, and 2.0\,pc respectively.  Columns 15--20
list the nominal parallax ($\parallax$), proper motion ($\propm$), and radial velocity ($\vr$) along with their standard errors. 
Objects with potentially problematic data have not been excluded.
The online table at CDS includes all objects with $\dphmed<$10\,pc (in a handful of cases for $\dphmed>2$\,pc the ID is Hipparcos, not Tycho).
% this is for 13 stars, even though 8 of them in fact have Tycho2 IDs. This is because the Gaia archive catalogue public.tycho fails
% to list TychoIDs for all Hipparcos stars.
\label{tab:periStats}}
\tabcolsep=0.14cm
\begin{tabular}{*{20}{r}}
\toprule
1 & 2 & 3 & 4 & 5 & 6 & 7 & 8 & 9 & 10 & 11 & 12 & 13 & 14 & 15 & 16 & 17 & 18 & 19 & 20 \\
\midrule
ID & & \multicolumn{3}{c}{$\tph$ / kyr} &  \multicolumn{3}{c}{$\dph$ / pc} &  \multicolumn{3}{c}{$\vph$ / \kms} & \multicolumn{3}{c}{\% samples} & $\parallax$ & $\sigma(\parallax)$ & $\propm$ & $\sigma(\propm)$ & $\vr$ & $\sigma(\vr)$ \\
          & & med & 5\% & 95\% & med & 5\% & 95\% & med & 5\% & 95\% & 0.5 & 1 & 2 & \multicolumn{2}{c}{\mas} & \multicolumn{2}{c}{\maspyr} & \multicolumn{2}{c}{\kms} \\
\midrule
\input{figures/encounters_2pc.tex}
\bottomrule
\end{tabular}
}
\end{table*}

\begin{figure}
\begin{center}
\includegraphics[width=0.5\textwidth, angle=0]{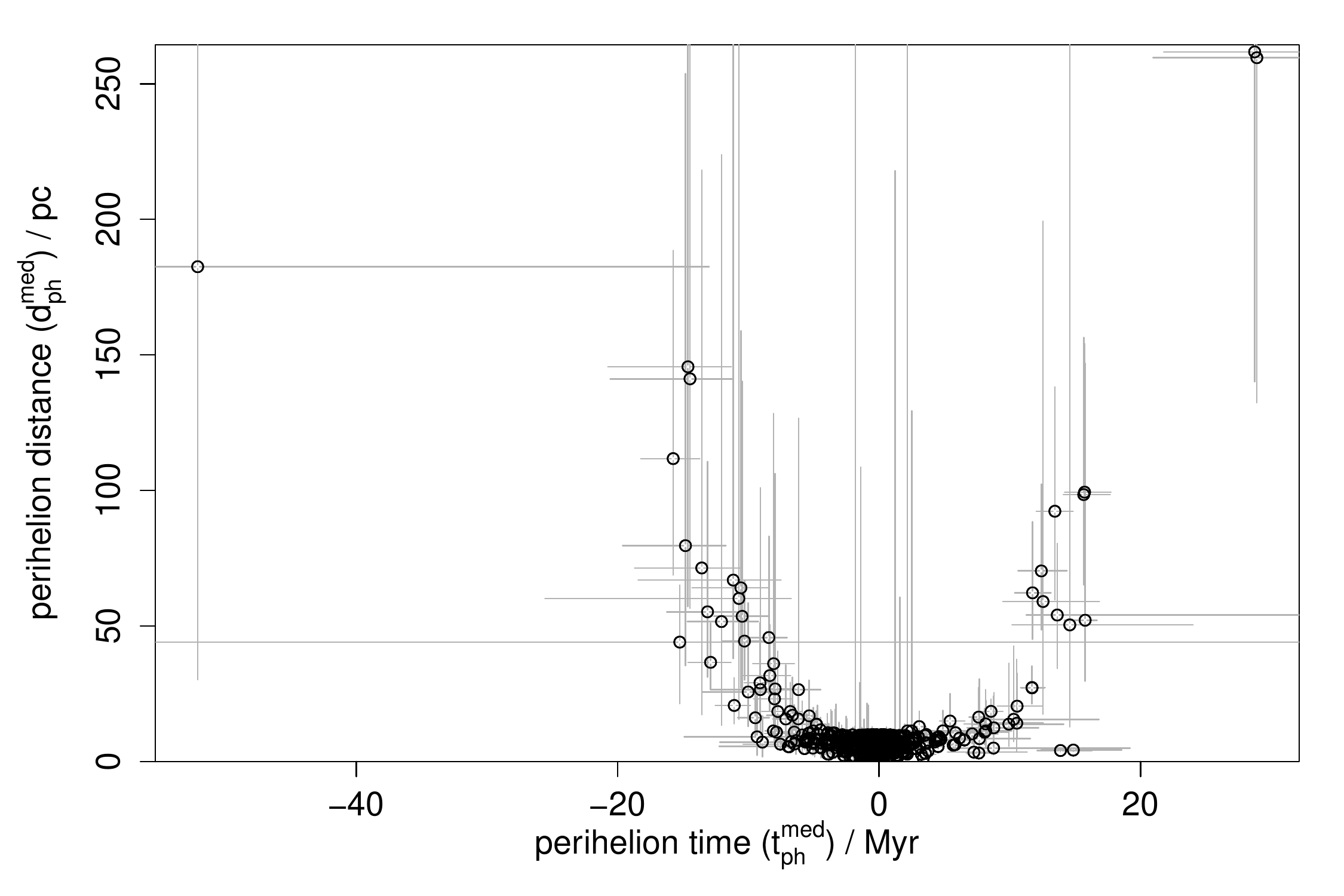}
\caption{Perihelion times and distances computed by orbit integration for all 725 objects. Open circles show the median of the perihelion time and distance distributions. The error bars show the limits of the 5\% and 95\% percentiles (which together form an asymmetric 90\% confidence interval). Note that this set of objects includes duplicates (see Table \ref{tab:dphSummary}).
\label{fig:dph_vs_tph_full_witherrors}}
\end{center}
\end{figure}

\begin{figure}
\begin{center}
\includegraphics[width=0.5\textwidth, angle=0]{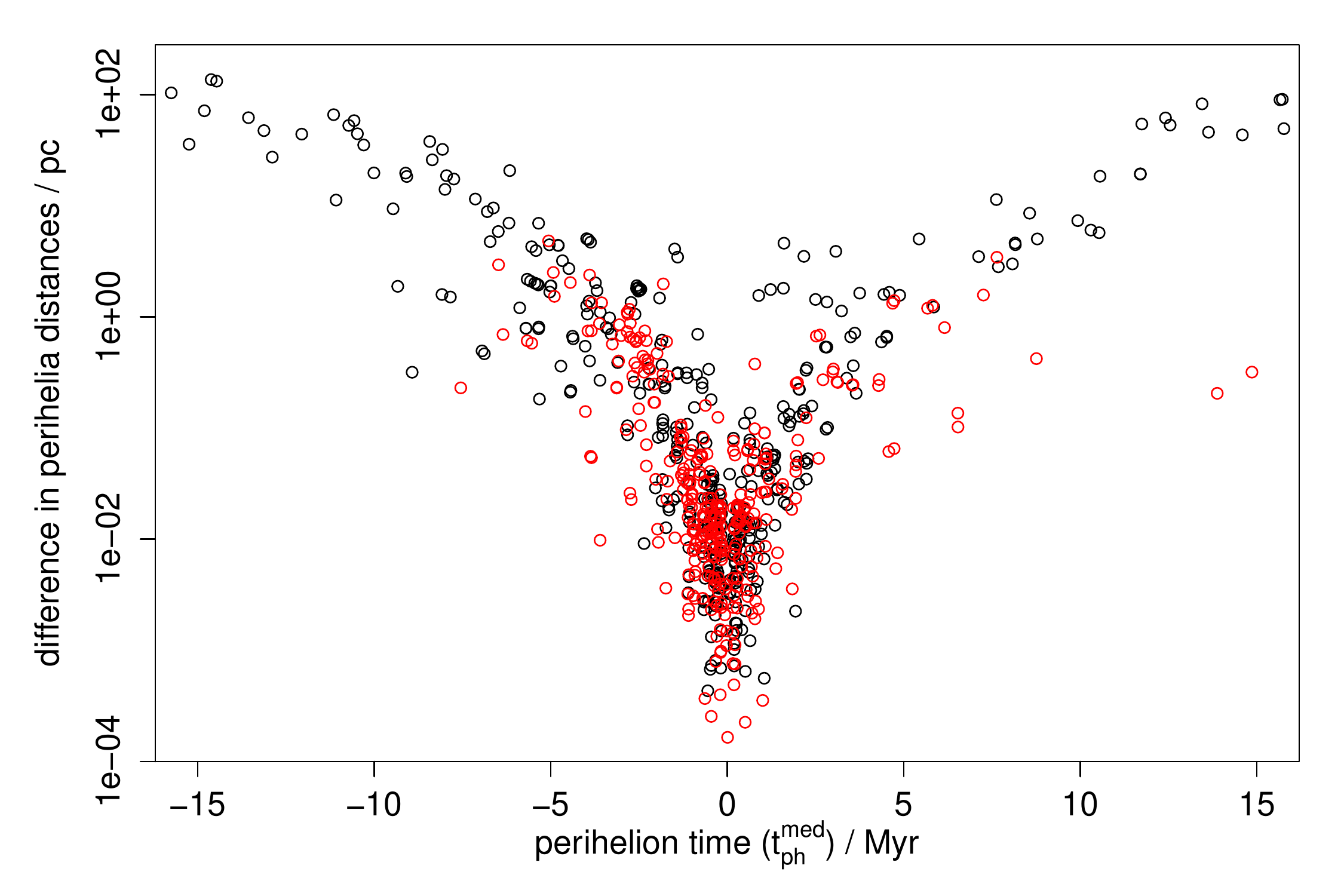}
\caption{Difference between perihelion estimates from orbit integration of surrogates ($\dphmed$) and the linear motion approximation ($\dphlin$), on a log scale. The black symbols are for $\dphmed>\dphlin$, the red symbols for $\dphmed<\dphlin$. The set of objects is the same as shown in 
Figure \ref{fig:dph_vs_tph_full_witherrors}, but excluding, for plotting purposes, the three points with $|\tphmed|>20$\,Myr. 
\label{fig:dph_minus_dphlin_vs_tph}}
\end{center}
\end{figure}

Figure \ref{fig:dph_vs_tph_full_witherrors} plots the perihelion times and distances. Although the objects were selected by the LMA to come within 10\,pc of the Sun, some of the orbit integrations result in much larger median perihelion distances. 
The differences between the two estimates -- LMA and median of orbit-integrated surrogates -- are shown in Figure \ref{fig:dph_minus_dphlin_vs_tph}. 
The differences arise both due to gravity and to the resampling of the data.
The largest differences are seen at larger (absolute) times. These objects have
small and low-significance parallaxes and proper motions. The resulting distribution over the perihelion parameters is therefore very broad (as seen from the large confidence intervals in Figure \ref{fig:dph_vs_tph_full_witherrors}), with the consequence that the median can differ greatly from the LMA value. If I use the LMA to find the perihelion of the surrogates and then compute their median, there are still large differences compared to the orbit integration median (up to 170\,pc; the plot of differences looks rather similar to Figure \ref{fig:dph_minus_dphlin_vs_tph}). This proves that the inclusion of gravity, and not just the resampling of the data, has a significant impact on the estimated perihelion parameters.
% This is best seen comparing the following plots in orbital_integration:
% results02/dph_minus_dphlin_vs_tph.pdf (which is shown in the present paper) with results02lma/dphorbit_minus_dphlma_vs_tphorbit_2.pdf
% - the latter shows the differences between two estimates of the median perihelion distance over as set of surrogates,
% one from orbit-computed perihelia (i.e. dphmed in the present paper), and the other LMA-computed perihelia 
% (which is computed by sample_multilma.R).
% You can also compare results02/dph_vs_tph_full_witherror.pdf (shown in the present paper) with results02lma/dph_vs_tph_full_witherror.pdf

\begin{figure}
\begin{center}
\includegraphics[width=0.5\textwidth, angle=0]{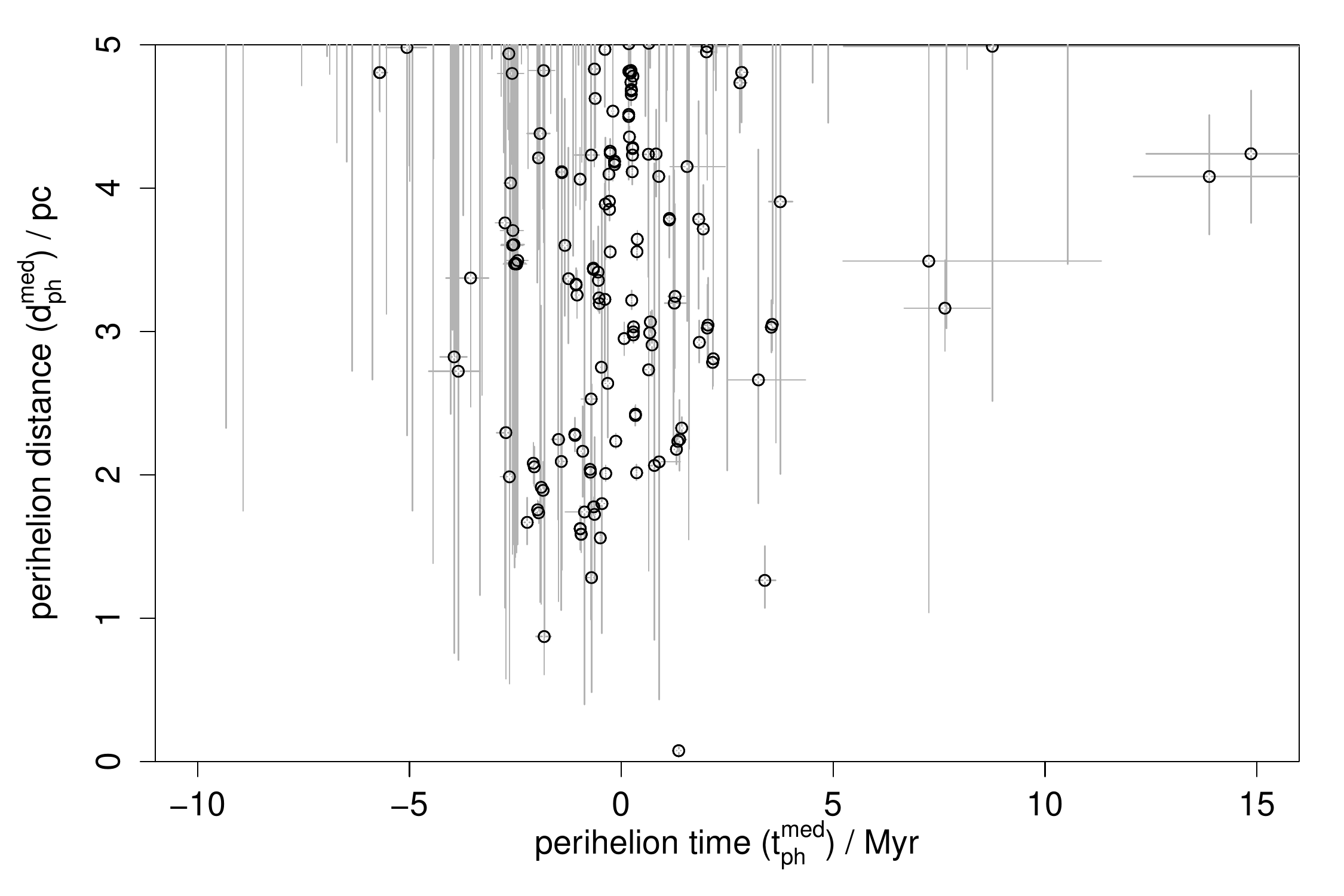}
\caption{As Figure \ref{fig:dph_vs_tph_full_witherrors}, but zoomed in to the range $0 \leq \dphmed \leq 5$.
The time axis is scaled to show all encounters in this distance range.
\label{fig:dph_vs_tph_0to5pc_witherrors}}
\end{center}
\end{figure}

\begin{figure}
\begin{center}
\includegraphics[width=0.5\textwidth, angle=0]{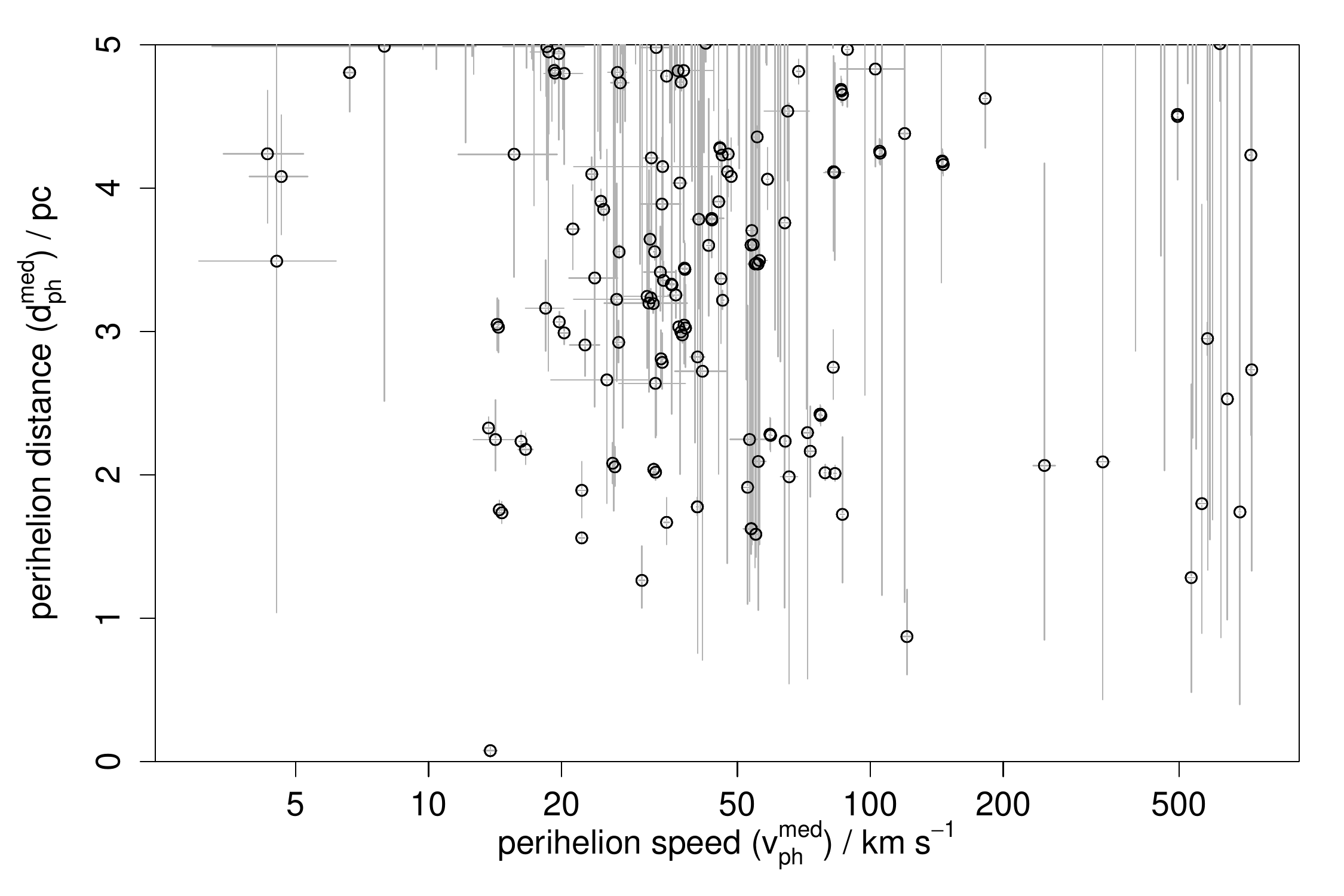}
\caption{Median perihelion velocities from the orbit integrations for those encounters shown in Figure \ref{fig:dph_vs_tph_0to5pc_witherrors}.
\label{fig:dph_vs_vph_0to5pc_witherrors}}
\end{center}
\end{figure}

As we are only interested in encounters which come within a few pc, Figure \ref{fig:dph_vs_tph_0to5pc_witherrors} zooms in to the perihelion range 0--5\,pc. The perihelion speeds for these objects 
are shown in Figure \ref{fig:dph_vs_vph_0to5pc_witherrors}.

Immediately apparent in Figure \ref{fig:dph_vs_tph_0to5pc_witherrors} is the very close encounter at 1.3\,Myr in the future. This is the K7 dwarf \object{Gl 710} (\object{Tyc 5102-100-1}, \object{Hip 89825}), known from many previous studies to be a very close encounter. In paper~1 I found a median 
encounter distance of 0.26\,pc (90\% CI 0.10--0.44\,pc).
TGAS reveals a much smaller proper motion than Hipparcos-2: $0.50 \pm 0.17$\,\maspyr\ as opposed to 
$1.8 \pm 1.2$\,\maspyr\ (the parallax is the same to within 2\%). Using the same radial velocity, my orbit integration now gives a median perihelion distance of 0.08\,pc (90\% CI 0.05--0.10\,pc), equivalently 
16\,000\,AU (90\% CI 10\,000--21\,000\,AU). This makes Gl\,710 once again the closest known stellar encounter.\footnote{Even at this close distance the gravitational interaction with the Sun can be neglected. Using the formulae in paper~1 (section 5.3), 
% hipparcos_paper1/deflection.R with vinf=13.8 km/s (as vt is only 0.05 km/s)
the path would be deflected by only 0.05\,\deg, bringing Gl\,710 just 7\,AU closer.} \cite{2016A&A...595L..10B} found a very similar value using the same data and a similar method. Although this close approach will take Gl\,710 well within the Oort cloud -- and its relative velocity at encounter is low (14\,\kms) -- its low mass (around 0.6\,\Msol) limits its perturbing influence. This is analysed by \cite{2016A&A...595L..10B}.

The second closest encounter is \object{Tyc 4744-1394-1} at $\dphmed$\,=\,0.87\,pc, two million years ago.
This is based on a RAVE radial velocity of $120.7 \pm 1.6$\,\kms. A second RAVE measurement gives $15.3 \pm 1.2$\,\kms, which puts the encounter at a much larger distance of $\dphmed=36.6$\,pc, 12.9\,Myr in the past.
% This is in encounters_Inf.dat
This discrepancy may suggest both measurements are unreliable.

\object{Tyc 1041-996-1} (\object{Hip 94512}) was found in paper~1 to encounter between 0.59--3.30\,pc or 
0.58--4.60\,pc (90\% CI), depending on the whether the Tycho-2 or Hipparcos-2 proper motion was adopted. TGAS puts a much tighter constraint on its perihelion distance, using the same radial velocity.

\object{Tyc 5033-879-1} is not in Hipparcos so was not in paper~1, and has not previously been reported as a close encounter. Its large radial velocity is from RAVE (DR5). The spectrum has a very low SNR, and the standard deviation computed by resampling is very high ({\tt StdDev\_HRV} = 246\,kms), so this measurement is probably spurious. This catalogue lists a second value 
of $-6.5 \pm 7.0$\,\kms\ for the same RAVE ID 
(\object{RAVE J160748.3-012060}). Another object with a different RAVE ID (\object{RAVE J160748.3-012059}) -- but very close by, and matched to the same Tycho ID by RAVE -- 
has $-11.6 \pm 2.1$\,\kms.

\object{Tyc 709-63-1} is \object{Hip 26335}. I found this to be a close encounter with very similar perihelion parameters in paper~1. The TGAS measurements are very similar to those from Hipparcos-2, but more precise.

I make no special provision for unresolved physical binaries in my search for encounters. The two objects \object{Tyc 4753-1892-1} and \object{Tyc 4753-1892-2} listed in Table \ref{tab:periStats} are in fact the two components of the spectroscopic binary \object{Hip 25240}, for which there is just one entry in TGAS (but both were listed separately in two RV catalogues). Their mutual centre-of-mass motion means the encounter parameters computed will be slightly in error. This could be corrected for in some known cases, but the many more cases of unidentified binarity would remain uncorrected. The situation will improve to some degree once Gaia includes higher order astrometric solutions for physical binaries (see section \ref{sec:future}).

Many of the other objects found in this study were likewise found to be close encounters in paper~1, sometimes with different perihelion parameters. The main reasons why some were not found in the earlier study are: not in the Hipparcos catalogue (for example because they were too faint); no radial velocity available; the TGAS and Hipparcos astrometry differ so much that $\dph$ was too large using the Hipparcos data to be reported in paper~1.

A few of the other closest encounters in Table~\ref{tab:periStats} have suspiciously high radial velocities. Some or all of these may be errors, although it must be appreciated that close encounters are generally those stars which have radial velocities much larger than their transverse velocities.  \object{Tyc 5383-187-1} is, according to Simbad, a B9e star. If the radial velocity from APOGEE (\object{SDSS 2M06525305-1000270}) is correct -- and this is questionable -- its perihelion speed was a whopping 687\,\kms.
% http://skyserver.sdss.org/DR13//en/tools/explore/summary.aspx?apid=apogee.apo25m.s.stars.4581.2M06525305-1000270
As its proper motion is very uncertain ($0.55 \pm 0.57$)\,\maspyr, the range of perihelion distances is huge. 
%There is even a 5\% chance that it passed within 0.4\,pc.  
The radial velocity for \object{Tyc 6975-656-1} is from RAVE. The value {\tt StdDev\_HRV} = 545\,\kms\ in that catalogue indicates its radial velocity, and therefore its perihelion parameters, are highly unreliable.

Slightly surprising is the fact that 18 of the 20 objects with $\dphmed<2$\,pc have encounters in the past. However, these 20 objects only correspond to 15 systems, and 13 of 15 is not so improbable. Moreover, for $\dphmed<3$\,pc the asymmetry is much weaker: 35 objects have encounters in the past vs 21 in the future.

\begin{figure}
\begin{center}
\includegraphics[width=0.5\textwidth, angle=0]{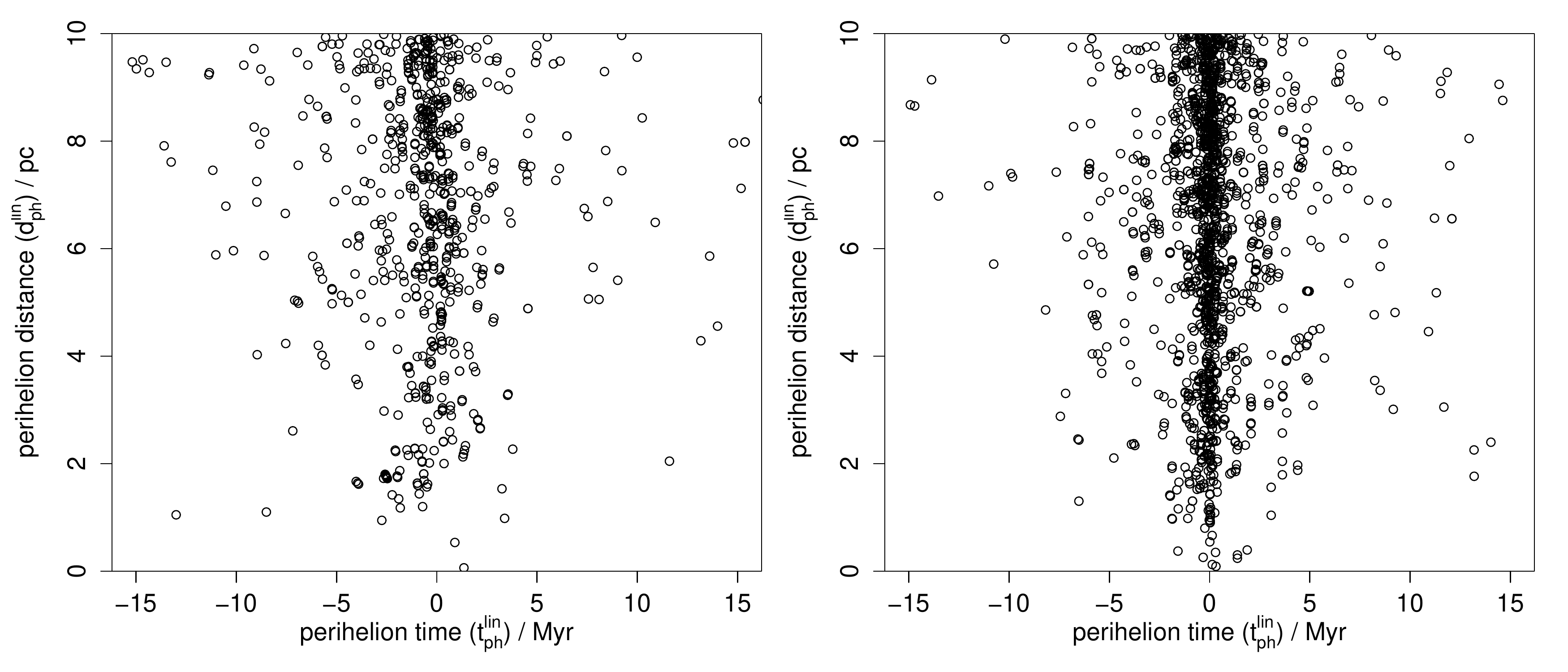}
\caption{Comparison of object encounters found in the present work with TGAS (left) and those found in paper~1 from Hipparcos (right). Both sets of estimates are from the linear motion approximation of the nominal data. In the left and right panels there are 16 objects and 1 object respectively which lie at times beyond the limits of the plots. Both panels shows all objects, not unique stars, so include duplicates.
% length(which(abs(1e-6*astromdat[,"tph"])>15)) = 16
%length(which(abs(1e-3*paper1LMA[,2])>16)) = 1
\label{fig:dphlin_vs_tphlin_TGAS_and_paper1}}
\end{center}
\end{figure}

Figure \ref{fig:dphlin_vs_tphlin_TGAS_and_paper1} compares the encounters found in paper~1 with those found in the present study. For this comparison I use the perihelion parameters computed by the LMA with the nominal data, as this permits an unbiased comparison out to 10\,pc. The most striking feature is that the Hipparcos-based study finds more encounters
 (1704 vs.\ 725 objects; 813 vs.\ 468 stars), despite the fact that the number of objects searched for encounters (i.e.\ those with complete and appropriate astrometry and radial velocities) 
is nearly four times larger in the present study: 397\,788 compared to 101\,363.
% section 2 of paper1 says 103555, from which non-positive parallaxes were removed.
% The lengths of the four input csv files in ~/cur/celestial_mechanics/stellar_encounters/hipparcos_paper1/linearmotion/
% in total (first number) and after removing non-positive parallaxes is
% xhip 46392 45207
% hip2gcs 12977 12953
% hip2pulkovo 35493 34816
% hip2rave4 8720 8387
% Total 103582 101363
% This difference is not due to the sigvr cut, as the figure 397788 is after applying that
Looking more closely, we also see that the present study finds comparatively few encounters at perihelion times near to zero. 
This is partly due to the fact that TGAS omits bright stars: brighter stars tend to be closer, and therefore near perihelion now (because the closer they are, the smaller the volume of space available for them to potentially come even closer).
TGAS's bright limit is, very roughly, G\,$=$\,4.5\,mag \citep{2016A&A...595A...2G}, whereas Hipparcos didn't have one.
For the encounters in paper~1 we indeed see a slight tendency for brighter stars to have smaller perihelion times.
% 134/162   = 0.83 with V<4.5 have |tphmean| < 1Myr
% 970/1542 = 0.63 with V>4.5 have |tphmean| < 1Myr
Some of the brightest stars in that study -- and missed by TGAS -- were also the closest encounters listed in Table~3 of paper~1, and include well-known stars such as \object{Alpha Centauri} and \object{Gamma Microscopii}. 
Yet as we shall see in section \ref{sec:completeness}), the encounter model
actually predicts a minimum in the density of encounters at exactly zero perihelion time.
Note that \object{Barnard's star} is missing because TGAS does not include the 19 Hipparcos stars with proper motions larger than 3500\,\mas\ \citep{2016A&A...595A...4L}.

In paper~1 I drew attention to the dubious data for a number of apparent close encounters. This included the closest encounter found in that study, \object{Hip 85605}. Unfortunately this and several other dubious cases are not in TGAS, so further investigation will have to await later Gaia releases. The closest-approaching questionable case from paper~1 in the present study is \object{Hip 91012} (\object{Tyc 5116-143-1}). This was flagged as dubious on account of its very high radial velocity in RAVE-DR4 ($-364 \pm 22$\,\kms; other RV catalogues gave lower values and thus more distant encounters). The TGAS astrometry agrees with Hipparcos, but the RAVE-DR5 radial velocity has a large uncertainty, and so was excluded by my new selection procedure. This same deselection applies to several other cases not found in the present study.
\cite{2015MNRAS.449.2459D} discuss in more detail some of the individual problematic cases in paper~1.

More interesting is \object{Hip 42525} (\object{Tyc 2985-982-1}).
The parallax is much smaller in TGAS than in Hipparcos-2: $5.9 \pm 0.5$\,\mas\ compared to $68.5 \pm 15.5$\,\mas. (The proper motions agree.) This puts its perihelion (according to the LMA) at around 90\,pc. In paper~1 I pointed out that the Hipparcos catalogue flagged this as a likely binary, with a
more likely parallax of $5.08 \pm 4.28$\,\mas. TGAS confirms this, but with ten times better precision. Its companion is \object{Tyc 2985-1600-1}, also in TGAS.

% Plot assuming 1/dph rather than 1/dph^2 is in results02/
\begin{figure}
\begin{center}
\includegraphics[width=0.5\textwidth, angle=0]{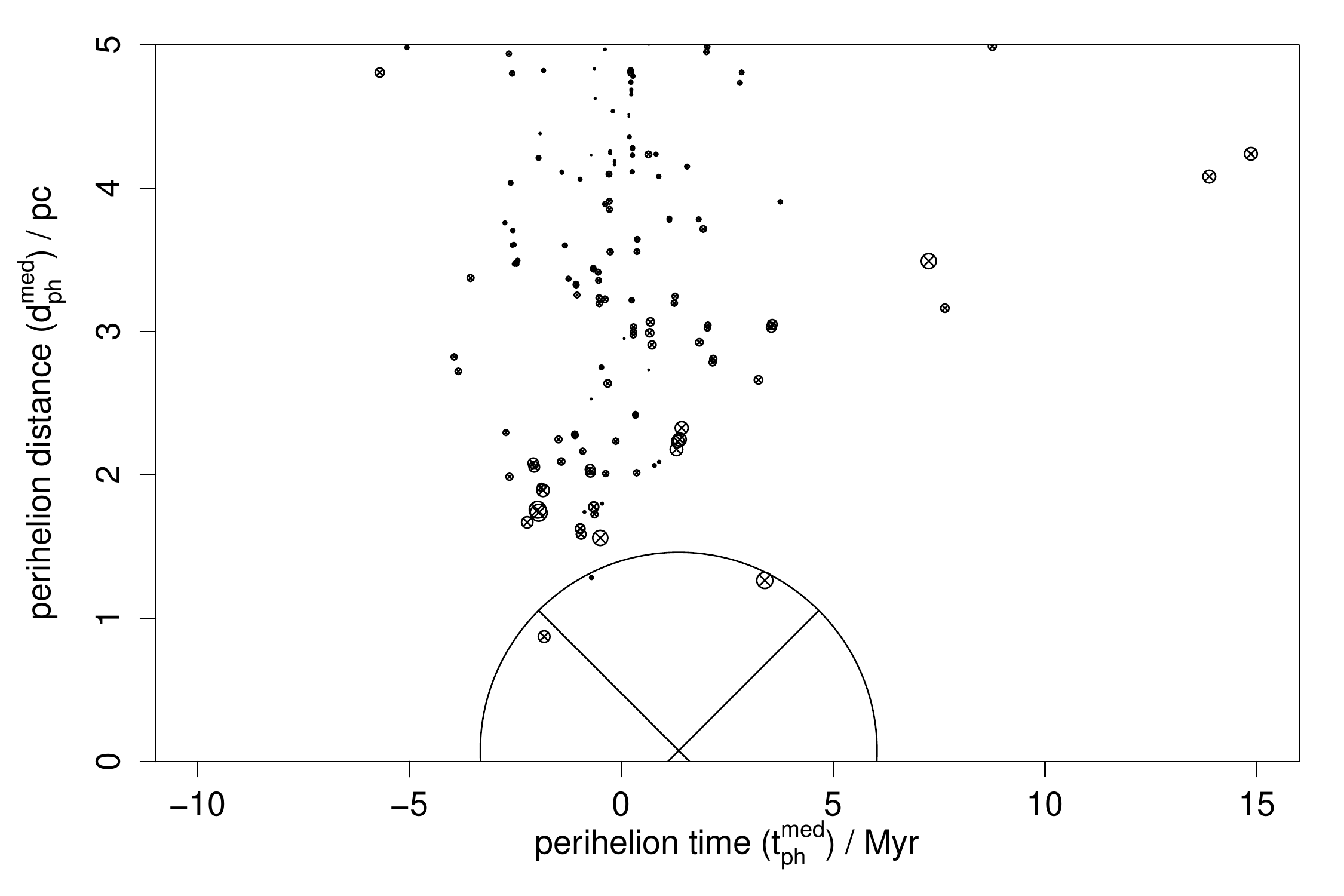}
\caption{As Figure \ref{fig:dph_vs_tph_0to5pc_witherrors}, but 
now plotting each object as a circle, the area of which is proportional to 
$1/(\vphmed (\dphmed)^2)$, which is proportional to the impulse transferred to the Oort cloud per unit mass of encountering star.
(The plotting symbol is a circle with a diagonal cross, but the latter is not apparent in the smallest symbols.)
The values for the plotted symbols range from $0.0088$ to $3.5$\,s\,km$^{-1}$\,pc$^{-2}$.
% sel <- which(periStats[,"dphMed"]<=5)
% range(sqrt(1/(periStats[sel,"dphMed"]^2*periStats[sel,"vphMed"])))
\label{fig:dph_vs_tph_impulse}}
\end{center}
\end{figure}

Equation \ref{eqn:impulse} approximates the change of momentum of an Oort cloud comet due to an encounter. As we do not (yet) know the masses of most of the encountering stars (not all have the required photometry/spectroscopy), this cannot be determined. But we can compute $1/(\vph \, \dph^2)$, which is proportional to the velocity change per unit mass of encountering star. This is visualized in Figure \ref{fig:dph_vs_tph_impulse} as the area of the circles. 
The importance of Gl\,710 compared to all other stars plotted is self-evident. Even if the second largest circle corresponded to a star with ten times the mass, Gl\,710 would still dominate the momentum transfer. This is in part due to the squared dependence on $\dph$. For very close encounters -- where the star comes within the Oort cloud -- the impulse is better described as $1/(\vph \, \dph)$ \citep{1976BAICz..27...92R}, in which case Gl\,710 is not so extreme compared to other stars. The impact of the different encounters is better determined by explicit numerical modelling, as done, for example, by \cite{2002A&A...396..283D}, \cite{2012P&SS...73..124R}, and \cite{2015MNRAS.454.3267F}.

\subsection{Comments on another TGAS study}\label{sec:otherstudy}

Just after I had submitted this paper for publication, there appeared on arXiv a paper by \cite{2017arXiv170608867B} reporting encounters found from TGAS/RAVE-DR5 by integrating orbits in a potential.

They find three encounters they consider as reliable, defined as having fractional errors of less than 10\% in initial position and velocity, and a radial velocity error of less than 15\,\kms (all three have $\sigvr<4.1$\,\kms). One of these is Gl\,710, for which they quote $\dph = 0.063 \pm 0.044$\,pc, broadly in agreement with my result. Their other two results are quite different from mine and are probably erroneous.

\begin{figure}
\begin{center}
\includegraphics[width=0.45\textwidth, angle=0]{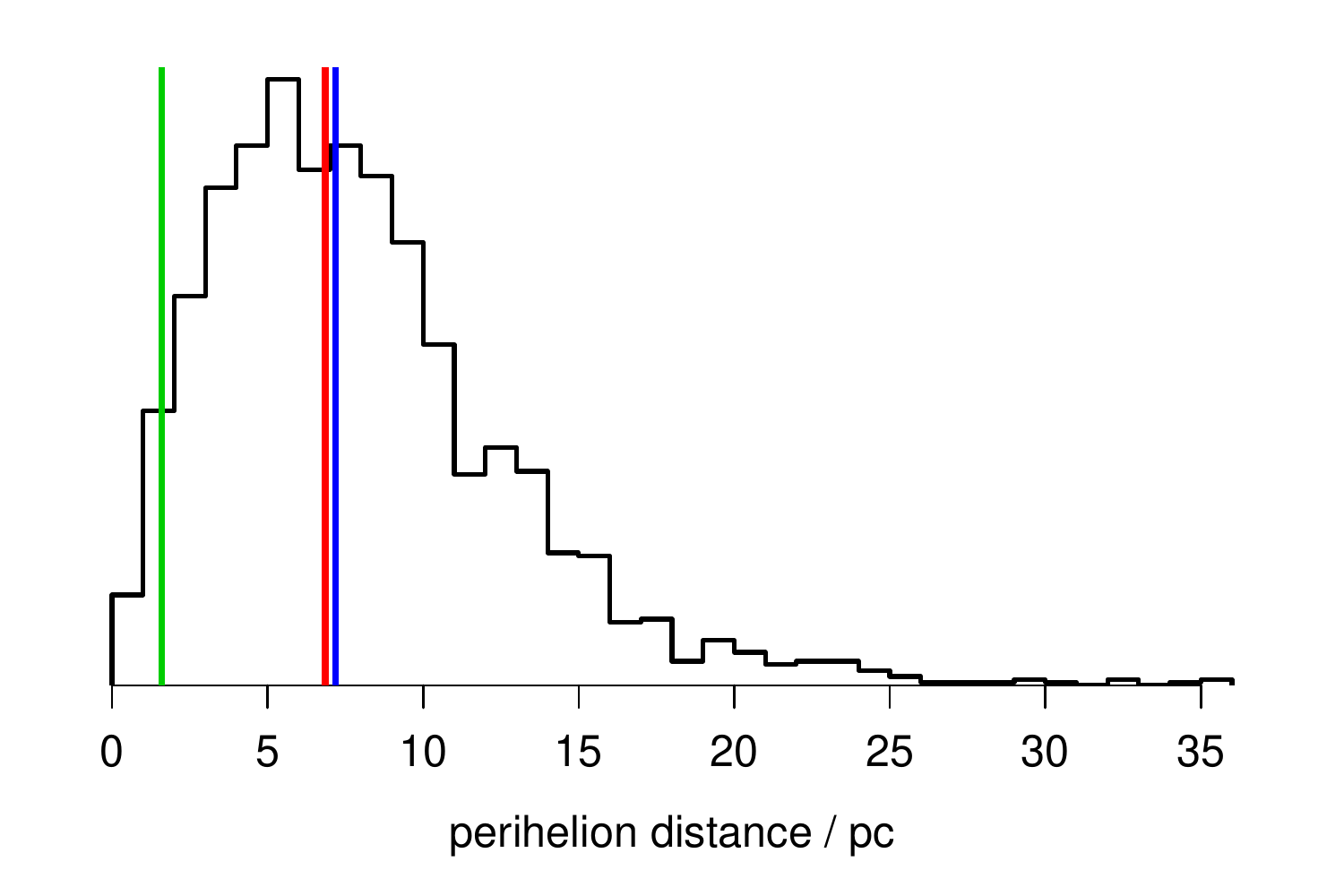}
\caption{Distribution of the perihelion distance (from the orbital integration of the 2000 surrogates) for 
\object{Tyc 6528-980-1}. The green, red, and blue lines (left to right) show the nominal, LMA, and median estimates. This demonstrates that the nominal estimate -- from integrating the orbit of just the nominal data -- can be very unrepresentative of the distribution we get when we account for the uncertainties in the data.
\label{fig:dphdist_Tyc_6528-980-1}}
\end{center}
\end{figure}

The first is \object{Tyc 6528-980-1}, which they put at $\dph = 0.86 \pm 5.6$\,pc. It's not clear from their brief description how this uncertainty is computed (seemingly the standard deviation of a set of surrogate orbits), but such a large symmetric uncertainty is inadmissible: just a 0.15-sigma deviation corresponds to an impossible negative perihelion distance.
%The distribution is intrinsically asymmetric and may not be summarized in this way. 
Their quoted best estimate of 0.86\,pc is also dubious. I derive a very different value {\em using the same input data}: $\dphmed=7.18$\,pc with a 90\% CI of  1.75--16.9\,pc (the large range being a result of the low significance proper motion). The difference is unlikely to arise from the different potential adopted. Even the LMA with the same data gives a perihelion distance of 6.9\,pc. 
I suspect their estimate comes directly from integrating an orbit for the nominal data. When I do this I get a perihelion distance of 1.60\,pc. Yet this value is not representative of the distribution of the surrogates, as can be seen in Figure~\ref{fig:dphdist_Tyc_6528-980-1}. 
Although the measured astrometric/RV data have a symmetric (Gaussian) uncertainty distribution, the nonlinear transformation
to the perihelion parameters means not only that this distribution is asymmetric, but also that the nominal data is not necessarily near the centre (it is particularly extreme in this case).
We have to correctly propagate the uncertainties not only to get a sensible confidence interval, but also to get a sensible point estimate.
The perihelion time I compute from the nominal data (-8.9\,Myr) is consistent with the median ($-8.9$\,Myr) and 90\% confidence interval ($-12.2$ to $-7.0$\,Myr) in this case, whereas Bobylev \& Bajkova report $-2.8 \pm 0.7$\,Myr.

Their other encounter (\object{Tyc 8088-631-1}) shows a similar problem. They quote a perihelion distance of $0.37 \pm 1.18$\,pc, whereas I find $\dphmed=1.2$\,pc with a 90\% CI of  0.6--5.9\,pc, and the LMA puts it at 0.95\,pc.

Bobylev \& Bajkova list in their Table 3 several other encounters which they consider less reliable.
These all have radial velocity uncertainties larger than 10\,\kms, so didn't enter my analysis.
Bobylev \& Bajkova do not mention several close several encounters which I find, even though their RAVE selection is more inclusive. This may be because they choose a smaller limit on perihelion distance.

%%%%%%%%%%%%%%%%%%%%%%%%%%%%%%%%%%%%%%%%%%%%%%%%%%%%%%%%%%%%
\section{Accounting for survey incompleteness}\label{sec:completeness}

% Plots in this section are made by stellar_encounters/completeness/make_nice_plots2.R

A search for stellar encounters is defined here as complete if it discovers all stars which come within a particular perihelion time and perihelion distance. Incompleteness arises when objects in this time/distance window are not {\em currently} observable. This may occur in a magnitude-limited survey because a star's large distance may render it invisible.  The completeness is quantified here by the function $C(\tph, \dph)$, which specifies the fraction of objects at a specified perihelion time and distance which are currently observable.

I estimate $C$ by constructing a model for encounters and determining their observability within a survey. The encounter data are not used. I first define a model for the distribution and kinematics of stars. In its most general form this is a six-dimensional function over position and velocity (or seven-dimensional function if we also include time evolution of the distributions). I then integrate the orbits of these stars to derive the probability density function (PDF) over perihelion time and distance, $\pmod(\tph, \dph)$. (Below I neglect gravity, so this integration will be replaced by the LMA.) The subscript ``mod'' indicates this has been derived from the model and, crucially, has nothing to do with the potential non-observability of stars.  I then repeat this procedure, but taking into account the selection function of the survey. In general this would involve a spatial footprint, an extinction map, as well as bright and faint magnitude limits and anything else which leads to stars not being observed. This allows us to derive $\pexp(\tph, \dph)$, where the ``exp'' subscript indicates this is the distribution we expect to observe in the survey (the asterisk indicates this is not a normalized PDF; more on this later). As shown in appendix \ref{appendix:completeness}, the ratio $\pexp(\tph, \dph)/\pmod(\tph, \dph)$ is the completeness function.

In principle we could extend the completeness function to also be a function of the perihelion speed, $\vph$ (and also the three other perihelion parameters which characterize directions). But I will assume we are 100\% complete in this parameter, i.e.\ we don't fail to observe stars on account of their kinematics.

\subsection{The model}\label{sec:themodel}

I implement the above approach using a very simple model for the encountering stars.
Let $r$ be the distance from the Sun to a star of absolute magnitude $M$ and space velocity (relative to the Sun) $\v$.
My assumptions are as follows:\footnote{``Homogeneous'' here means ``the same at all points in space''.}
\begin{enumerate}
\item the stellar spatial distribution is isotropic, described by $P(r)$. % but not homogeneous: density is lower further away!
\item the stellar velocity distribution is isotropic and homogeneous, described by $P(\v)$;
\item the stellar luminosity function, $P(M)$, is homogeneous;
\item there is no extinction;
\item stellar paths are linear (no gravity);
\item the survey selection function, $S(m)$, is only a function of the apparent magnitude $m$. It specifies the fraction of objects observed at a given magnitude. 
\end{enumerate}
Note that $P(r)$, $P(\v)$, and $P(M)$ are all one-dimensional functions, which will greatly simplify the computation of $\pmod(\tph, \dph)$ and $\pexp(\tph, \dph)$.
The symmetry in these assumptions means any distribution will be symmetric in time. We therefore only need to consider future encounters. 

Although my model is very simple (see the end of this subsection), it turns out to be reasonably robust to some changes in the assumed distributions, as we shall see. This is because we are only interested in how $\pmod(\tph, \dph)$ {\em changes} under the introduction of the selection function.  The poorly-defined selection functions of the TGAS data and the RV catalogues mean we would gain little from constructing a more complex model, yet we would pay a price in interpretability.

\begin{figure}
\begin{center}
\includegraphics[width=0.45\textwidth, angle=0]{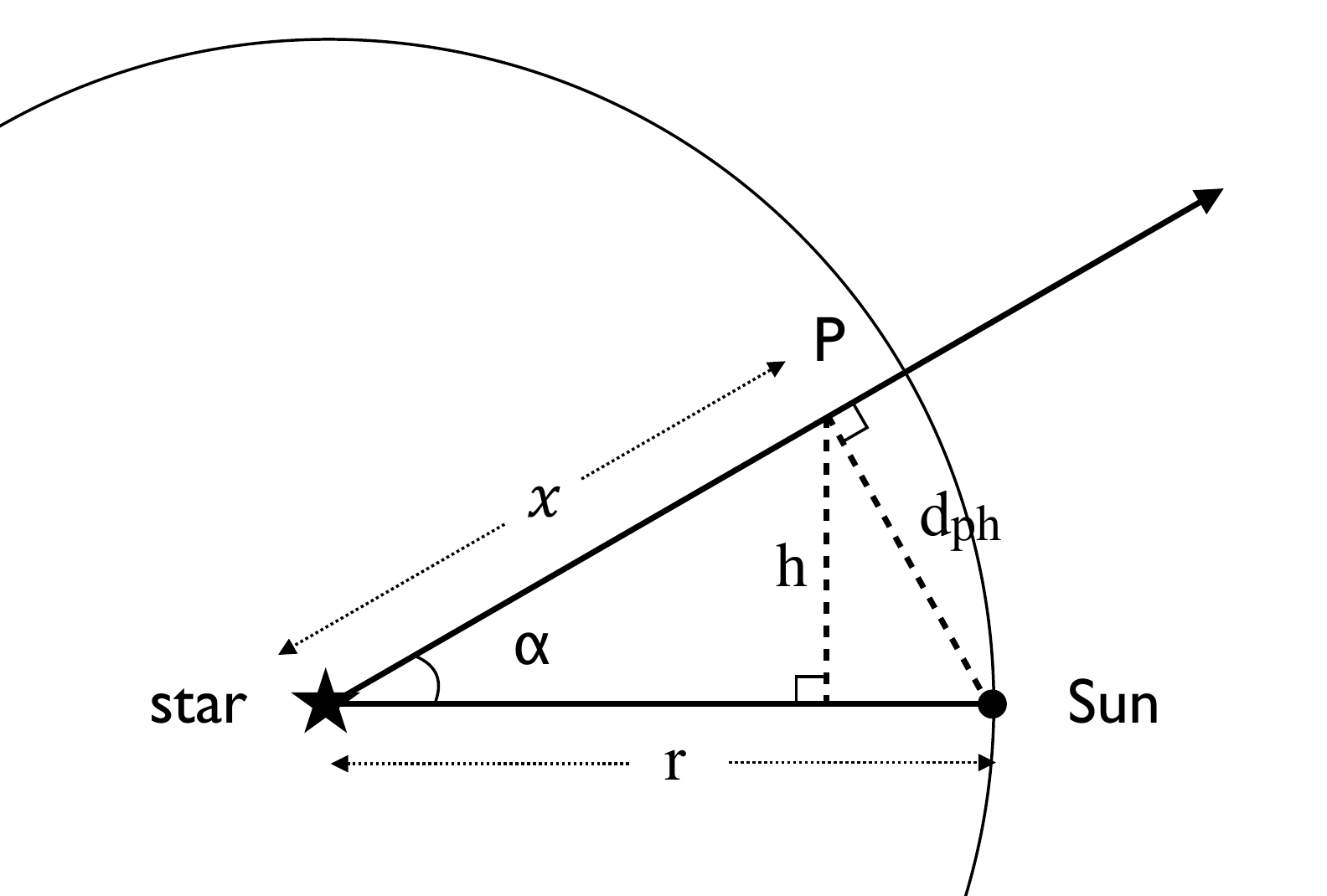}
\caption{The geometry of a close encounter for a star initially at distance $r$ and reaching a perihelion distance of $\dph$ at point $P$, whereby $\dph \leq r$.
\label{fig:completeness_geometry_cut}}
\end{center}
\end{figure}

Using these assumptions I now derive $P(\tph, \dph)$. The initial steps are valid for both the ``mod'' and ``exp'' distributions, so I omit the subscripts where they are not relevant.

\begin{figure}
\begin{center}
\includegraphics[width=0.5\textwidth, angle=0]{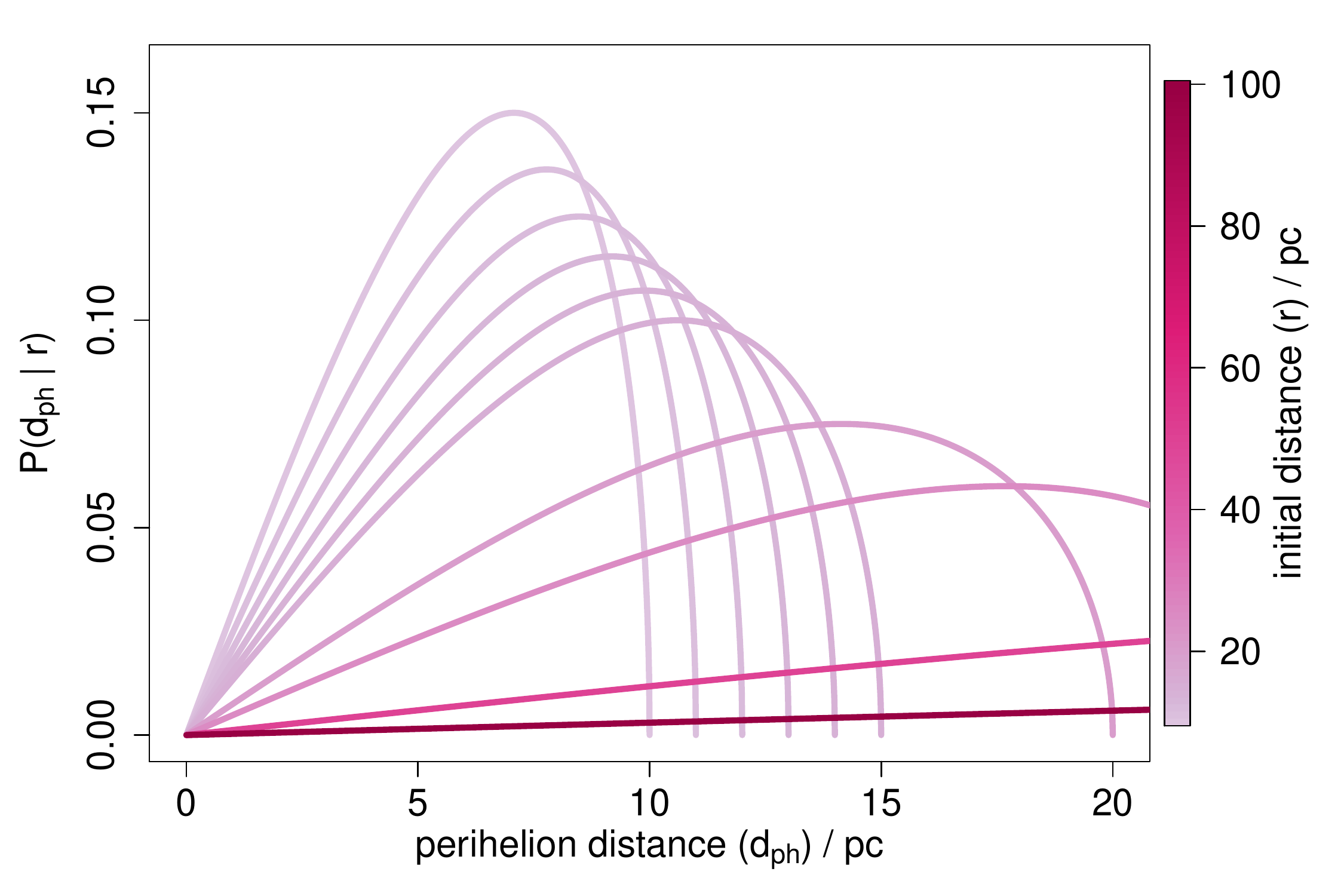}
\caption{The PDF over the perihelion distance for stars with different initial distances $r=10, 11, 12, 13, 14, 15, 20, 25, 50, 100$\,pc, shown in different colours, as described by the model (section \ref{sec:themodel}). The lines for $r>20$\,pc extend beyond the right boundary of the plot, but have the same shape, dropping to zero at $\dph=r$.
\label{fig:P_dph_given_r}}
\end{center}
\end{figure}

The geometry for an arbitrary star currently at distance $r$ from the Sun is shown in Figure \ref{fig:completeness_geometry_cut}.
The star travels a distance $x$ to reach perihelion at point $P$, when its distance from the Sun is $\dph$. Note that $\dph \leq r$, otherwise the star has already been at perihelion (and I neglect past encounters due to time symmetry). To achieve a perihelion between distance $\dph$ and $\dph + \delta\dph$, this star must pass through a circular ring
of inner radius $h$ and infinitesimal width $\delta\dph$. Note that
$\delta\dph$ is parallel to $\dph$ and so perpendicular to the direction of motion.
%(The axis of this ring is the line star--Sun in the diagram, so the sides of the ring are not parallel to its axis.)
The area of this circular ring as seen from the initial position of the star is therefore $\delta A = 2\pi h \delta\dph$, its circumference times its width.
As all directions of travel for the star are equally probable (isotropic velocity distribution), the probability that this star
% initially at distance $r$ from the Sun
has a perihelion distance between $\dph$ and $\dph + \delta\dph$ is just proportional to $\delta A$. That is
\begin{alignat}{2}
P(\dph \given r) \, \delta\dph \,&\propto\, \delta A \ ,  \hspace*{1em}{\rm so}\\
P(\dph \given r) \,&\propto\, h \label{eqn:h_over_Z} \ .
\end{alignat}
From the diagram we see that
\begin{equation}
\sin\alpha \,=\, \frac{\dph}{r} \,=\, \frac{h}{x}
\end{equation}
and
\begin{equation}
x^2 = r^2 - \dph^2
\label{eqn:x}
\end{equation}
from which it follows that
\begin{equation}
h \,=\, \dph \left(1 - \frac{\dph^2}{r^2}\right)^{1/2} \ .
\end{equation}
We can normalize analytically to get
\begin{equation}
P(\dph \given r) \,=\, \frac{3\dph}{r^2} \left(1 - \frac{\dph^2}{r^2}\right)^{1/2} \hspace*{1em}{\rm where}\hspace*{1em} \dph \leq r \ .
\end{equation}
Plots of this distribution for different values of $r$ are shown in Figure \ref{fig:P_dph_given_r}. 
This may be counter-intuitive. For {\em fixed} $r$ the PDF is proportional to $h$ (equation \ref{eqn:h_over_Z}). 
As we increase $\dph$ from zero, $h$ initially increases from zero. As the line Sun--P is always perpendicular to the line P--star,  increasing $\dph$ beyond some point leads to $h$ decreasing. The PDF therefore has a maximum at $\dph<r$. 

% Plot from completeness/model21

We now compute the distribution over perihelion times. As velocities are constant, the time of the encounter, $\tph$, is $c_0x/\v$, where $\v$ is the velocity. Using parsecs for distance, years for time, and \kms\ for velocity, $c_0 = 0.97779 \times 10^6$. For a given $x$ there is a one-to-one correspondence between $\v$ and $\tph$. As $x$ is determined by $\dph$ and $r$, this means we can write
\begin{equation}
P(\tph \given \dph, r) \, \deriv\tph \,=\ P(\v \given \dph, r) \, \deriv \v .
\end{equation}
Because we are assuming an isotropic and homogeneous velocity distribution, $P(\v)$, we can write the term on the right as $P(\v)\,\deriv\v$. It then follows that
\begin{alignat}{2}
P(\tph \given \dph, r) \,&=\ P(\v) \left| \frac{\deriv\v}{\deriv\tph} \right| \\
                                 \,&=\ P\left(\v=\frac{c_0 x}{\tph}\right) \frac{c_0 x}{\tph^2} \hspace*{1em}{\rm where}\hspace*{1em} \dph \leq r
\end{alignat}
with $x$ given by equation \ref{eqn:x}. This distribution is normalized (provided $P(\v)$ is).

The model PDF we want is related to the two PDFs we just derived by marginalization, namely
\begin{equation}
\pmod(\tph, \dph) \,=\ \int_{\dph}^{\infty} P(\tph \given \dph, r) P(\dph \given r) \pmod(r) \, \deriv r \ .
\end{equation}
Typically, this integral must be computed numerically. I do this below by integrating on a grid.

The impact of the selection function is to reduce the number of stars we would otherwise see.
The selection function depends only on apparent magnitude which, as I neglect extinction, is $m = M + 5(\log r - 1)$. 
Consequently the selection function only modifies the distance distribution, $\pmod(r)$, to become
\begin{equation}
\pexp(r) \,=\, \pmod(r) \int_{-\infty}^{+\infty} P(M)S(m) \, \deriv M \ .
\end{equation}
This PDF is not normalized (I use the asterisk to remind us of this). As $0 \leq S(m) \leq 1$, 
$\pexp(r) \leq \pmod(r)$ for all $r$, i.e.\ the change from $\pmod(r)$ to $\pexp(r)$ reflects the absolute decrease in the distribution due to the selection function.
This integral I also evaluate on a grid. The expected PDF over the perihelion time and distance is then
\begin{equation}
\pexp(\tph, \dph) \,=\ \int_{\dph}^{\infty} P(\tph \given \dph, r) P(\dph \given r) \pexp(r) \, \deriv r 
\end{equation}
which is also not normalized. 
The completeness function is {\em defined} as 
\begin{equation}
C(\tph, \dph) \,=\, \frac{\fexp(\tph, \dph)}{\fmod(\tph, \dph)} \ . 
\label{eqn:cdef1}
\end{equation}
where $\fmod(\tph, \dph)$ and $\fexp(\tph, \dph)$ are the model and expected encounter flux -- number of encounters per unit perihelion time and distance -- respectively. As shown in appendix \ref{appendix:completeness}, the completeness can be written as
\begin{equation}
C(\tph, \dph) \,=\, \frac{\pexp(\tph, \dph)}{\pmod(\tph, \dph)} \ ,
\label{eqn:completeness}
\end{equation}
which is the ratio of the two PDFs I just derived.
For all $\tph$ and $\dph$, $0 \leq C(\tph, \dph) \leq 1$. This completeness function gives the probability (not probability density!) of observing an object at a given perihelion time and distance. I will use it in section \ref{sec:encounter_frequency} to infer the encounter rate. 

% Plots from completeness/model21
\begin{figure}
\begin{center}
\includegraphics[width=0.24\textwidth, angle=0]{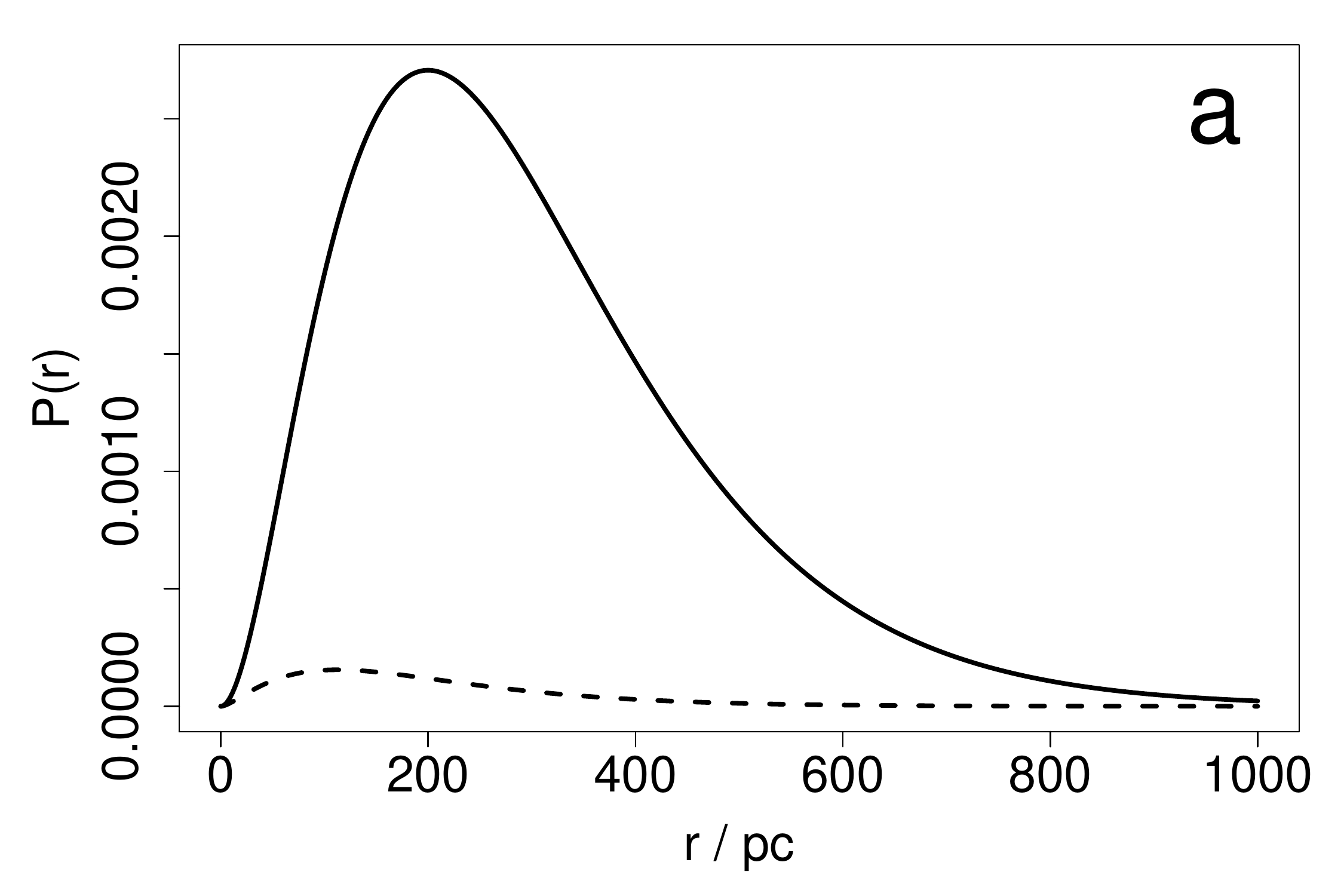}
\includegraphics[width=0.24\textwidth, angle=0]{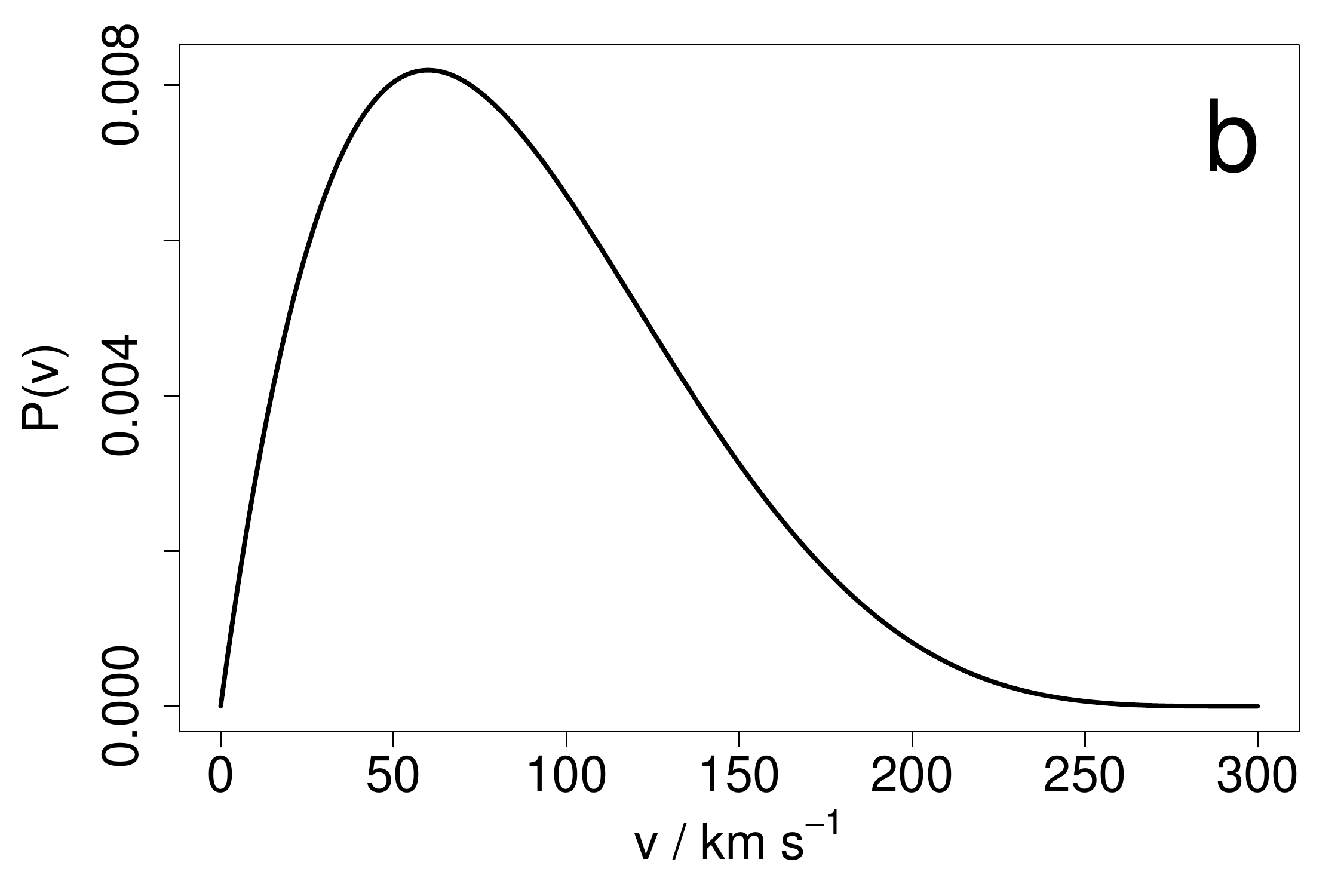}
\includegraphics[width=0.24\textwidth, angle=0]{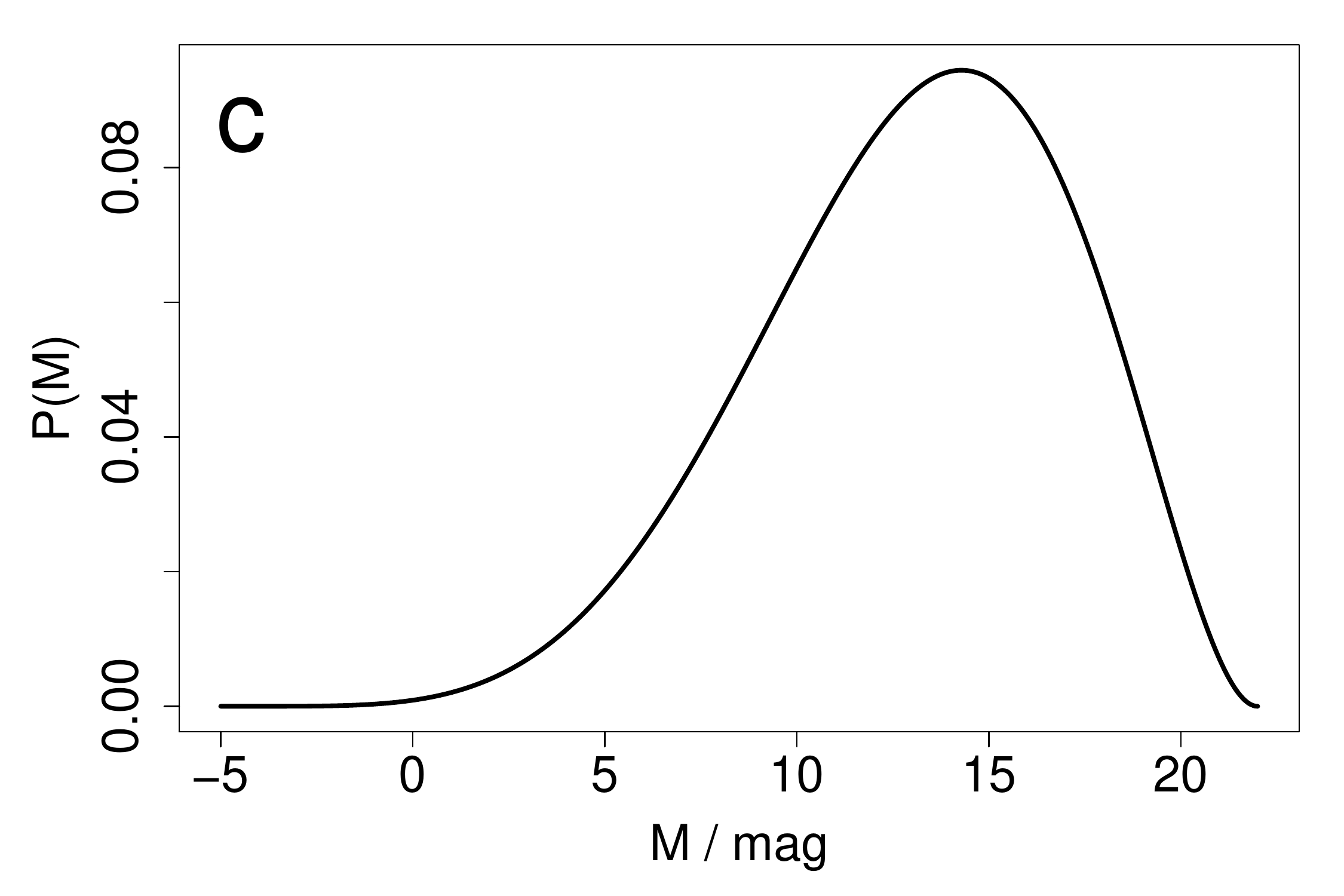}
\includegraphics[width=0.24\textwidth, angle=0]{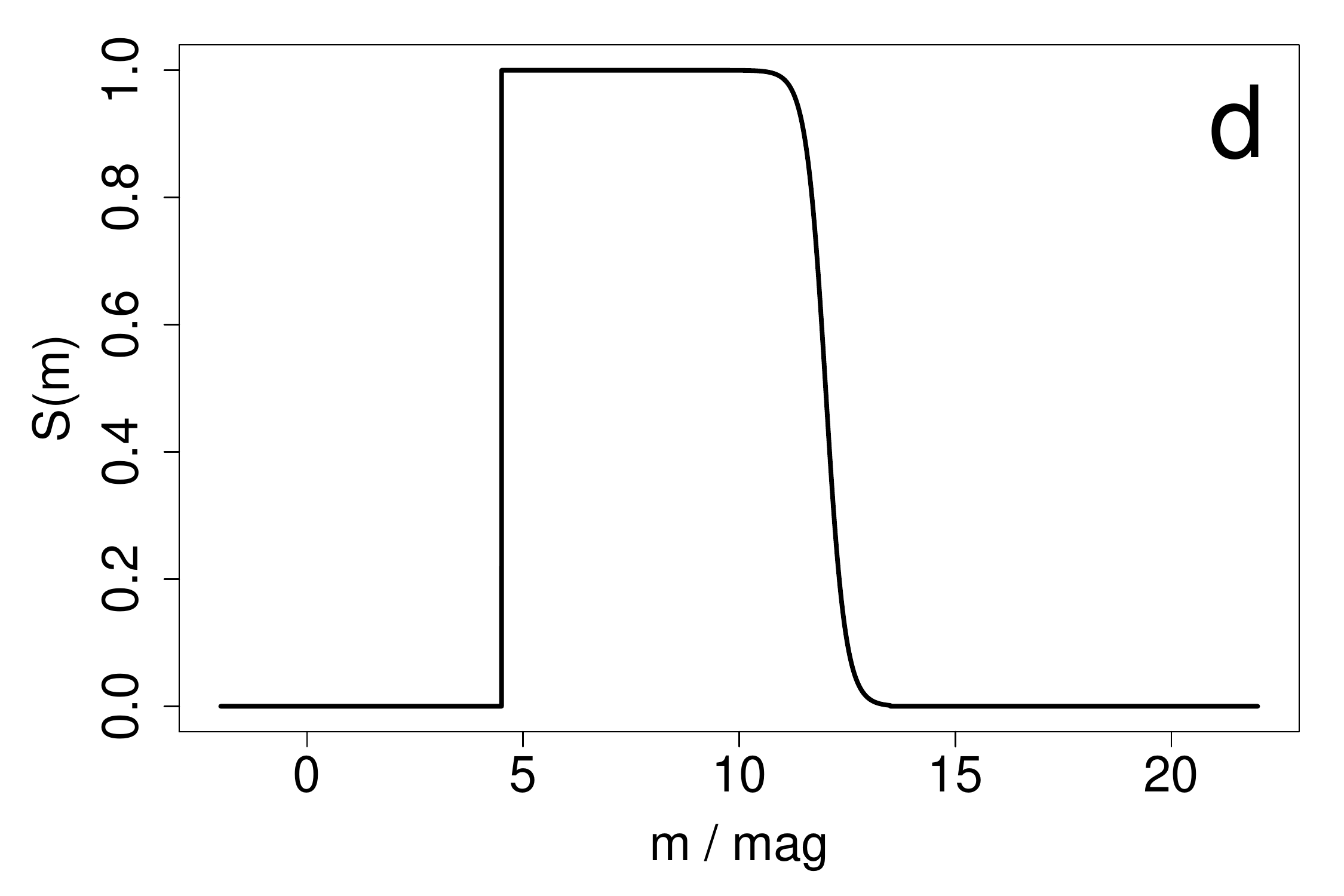}
\caption{The input distributions for the completeness model are the (a) distance (solid line), (b) space velocity, (c) absolute magnitude (normalized PDFs). The selection function (on apparent magnitude) is shown in panel (d). This modifies the model distance distribution to become the expected distribution shown as the dashed line in panel (a).
\label{fig:completeness_model_inputs}}
\end{center}
\end{figure}

To compute the completeness function we need to select forms for the various input distributions. These are shown in Figure \ref{fig:completeness_model_inputs}.
For the distance distribution I use an exponentially decreasing space density distribution \citep{2015PASP..127..994B}
\begin{equation}
\pmod(r) \,=\, \frac{1}{2\rlen^3}\,r^2e^{-r/\rlen}  \hspace*{1em}{\rm where}\hspace*{1em} r > 0
\label{eqn:pmod_r}
\end{equation}
and $\rlen$ is a length scale, here set to 100\,pc. This produces a uniform density for $r \ll \rlen$. 
This is shown in panel (a) as the solid line. 
This form has been chosen mostly for convenience, with a length scale which accommodates the decrease in density due to the disk scale height.
The dashed line in panel (a) is $\pexp(r)$. As expected, the selection function reduces the characteristic distance out to which stars can be seen. Panels (b) and (c) show the velocity and absolute magnitude distributions respectively. These are both implemented as shifted/scaled beta distributions, so are zero outside the ranges plotted. Again these have been chosen mostly for convenience. The luminosity distribution is a rough fit to the distribution in \cite{2015MNRAS.451..149J}, with a smooth extension to fainter magnitudes. The lower end of the velocity distribution is a reasonable fit to the TGAS stars, but I have extended it to larger velocities to accommodate halo stars. This distribution ignores the anisotropy of stellar motions near to the Sun \citep[e.g.][]{2014MNRAS.442.3653F}. But as I am not concerned with the $(l,b)$ distribution of encounters, this is not too serious (I effectively average over $(l,b)$).
% velocity: 
% dbeta(v/vmax, shape1=2, shape2=5)/vmax
% with vmax=300. Max at vmax*(shape1 - 1)/(shape1 + shape2 - 2) = 60.
% mag:
% dbeta((M-Mrange[1])/(Mrange[2]-Mrange[1]), shape1=6, shape2=3)/(Mrange[2]-Mrange[1])
% with Mrange=c(-5,22)
None of these distributions is particularly realistic, but are reasonable given the already limiting assumptions imposed at the beginning of this section. 
Earlier attempts at completeness correction for encounter studies used even simpler assumptions.
In section~\ref{sec:sensitivity} I investigate the sensitivity of the resulting completeness function to these choices.  Panel (d) models the TGAS selection function, as derived from the information given in \cite{2000A&A...355L..27H} and \cite{2016A&A...595A...4L}. Due to the preliminary nature of the TGAS astrometric reduction, the complex selection function of the Tycho catalogue, and what was included in TGAS, this is a simple approximation of the true (unknown) TGAS selection function. In practice the selection function does not depend only on apparent magnitude. In particular, there are some bright stars fainter than my bright limit of G=4.5 that are not in TGAS.  The only way to accommodate for such an incompleteness would be to inflate the encounter rate based on other studies, but this comes with its own complications, such as how to model {\em their} incompletenesses. The goal in the present work is to explore the consequences of a very simple model. Once we have data with better-defined selection functions, more complex -- and harder to interpret -- models will become appropriate and necessary.

\subsection{Model predictions}\label{sec:completeness_predictions}

% Plots from completeness/model21
\begin{figure*}
\begin{center}
\includegraphics[width=0.49\textwidth, angle=0]{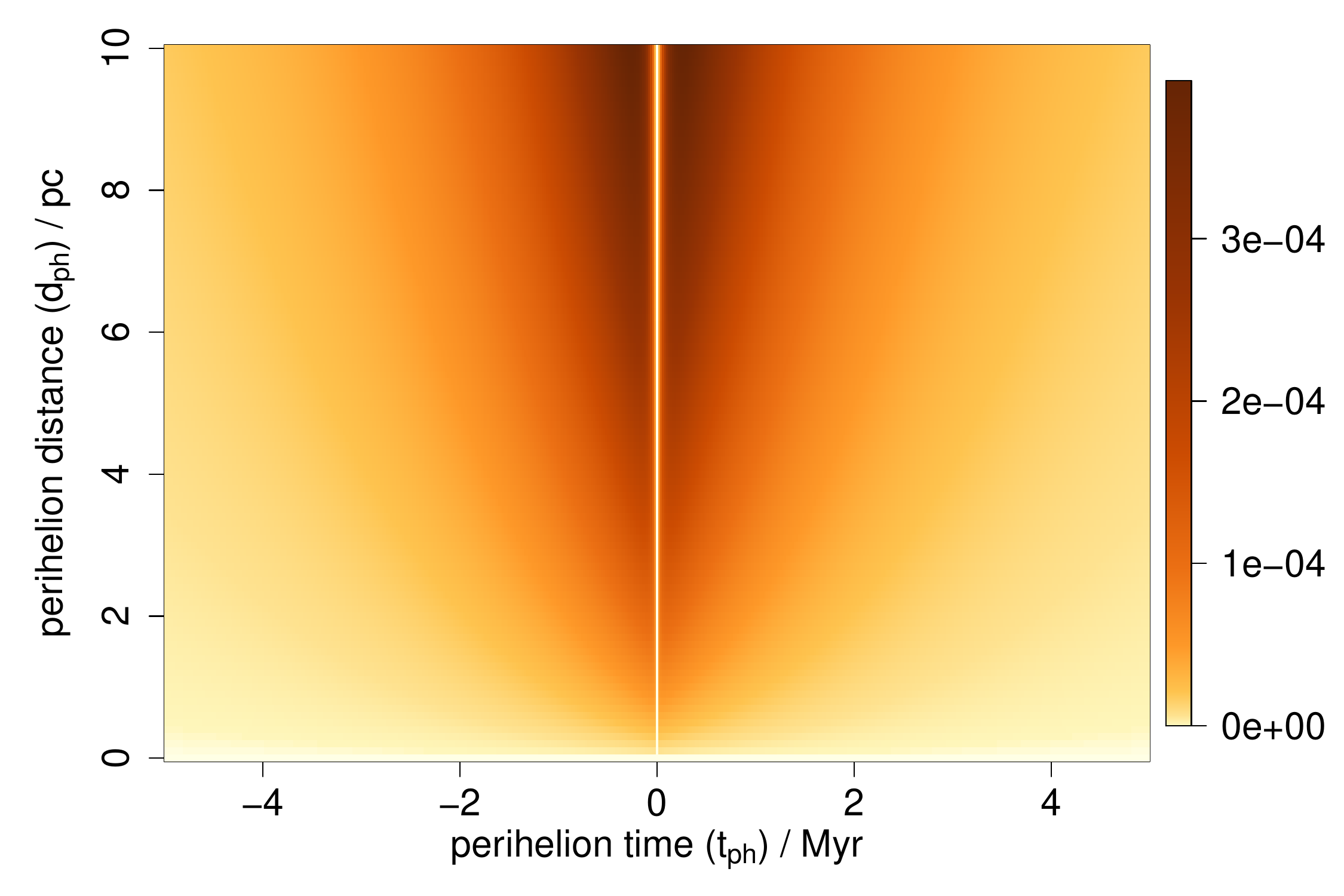}
\includegraphics[width=0.49\textwidth, angle=0]{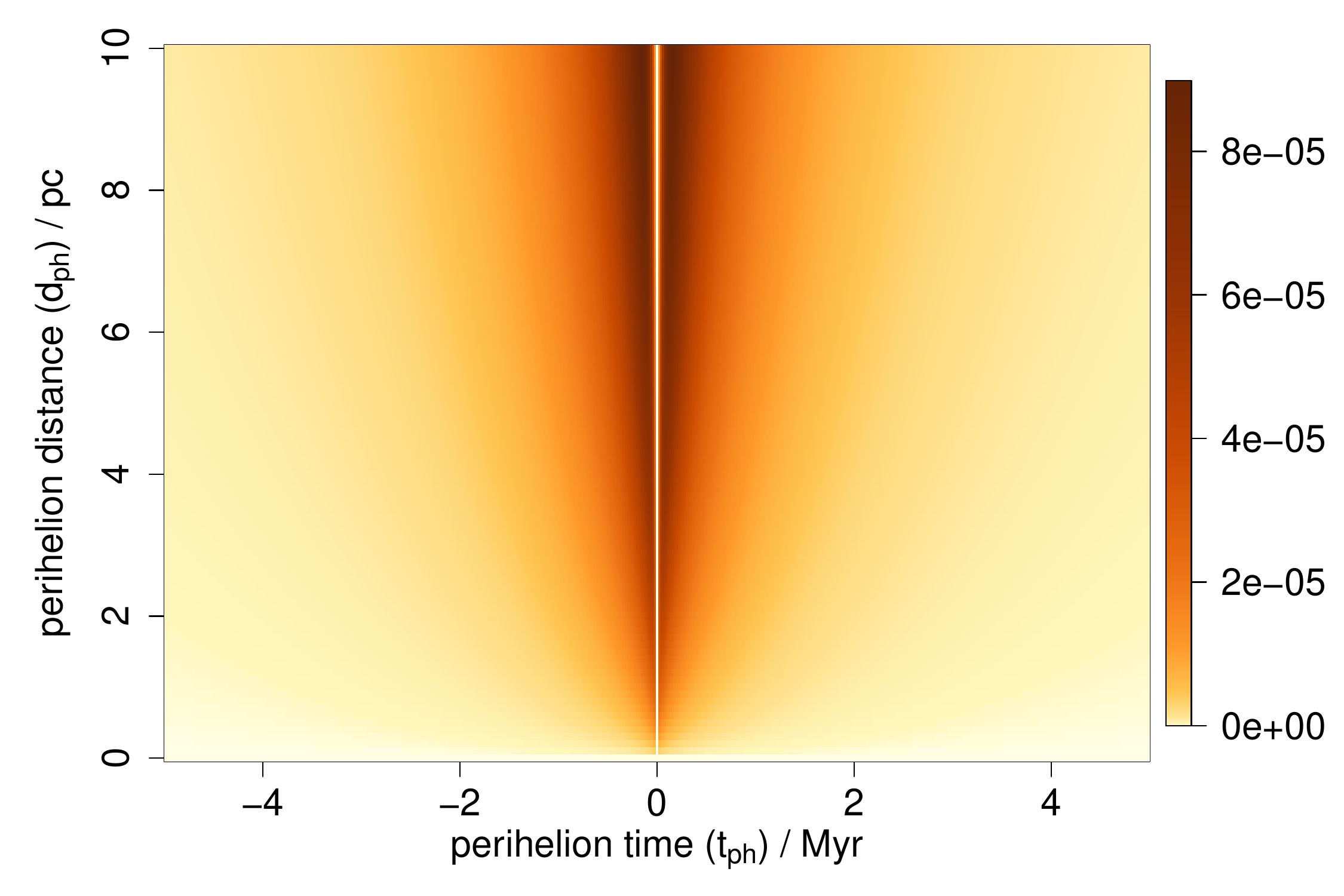}
\includegraphics[width=0.49\textwidth, angle=0]{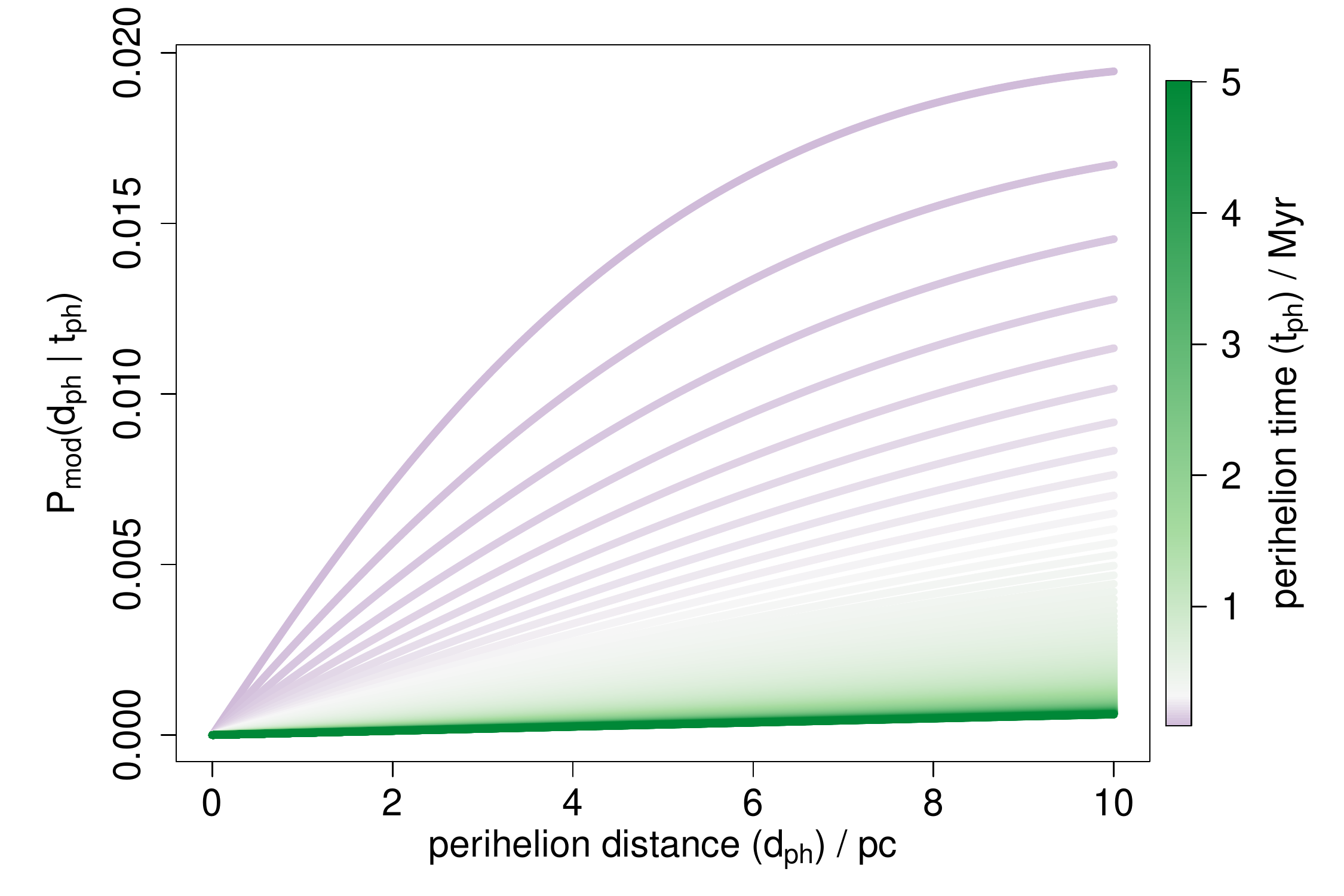}
\includegraphics[width=0.49\textwidth, angle=0]{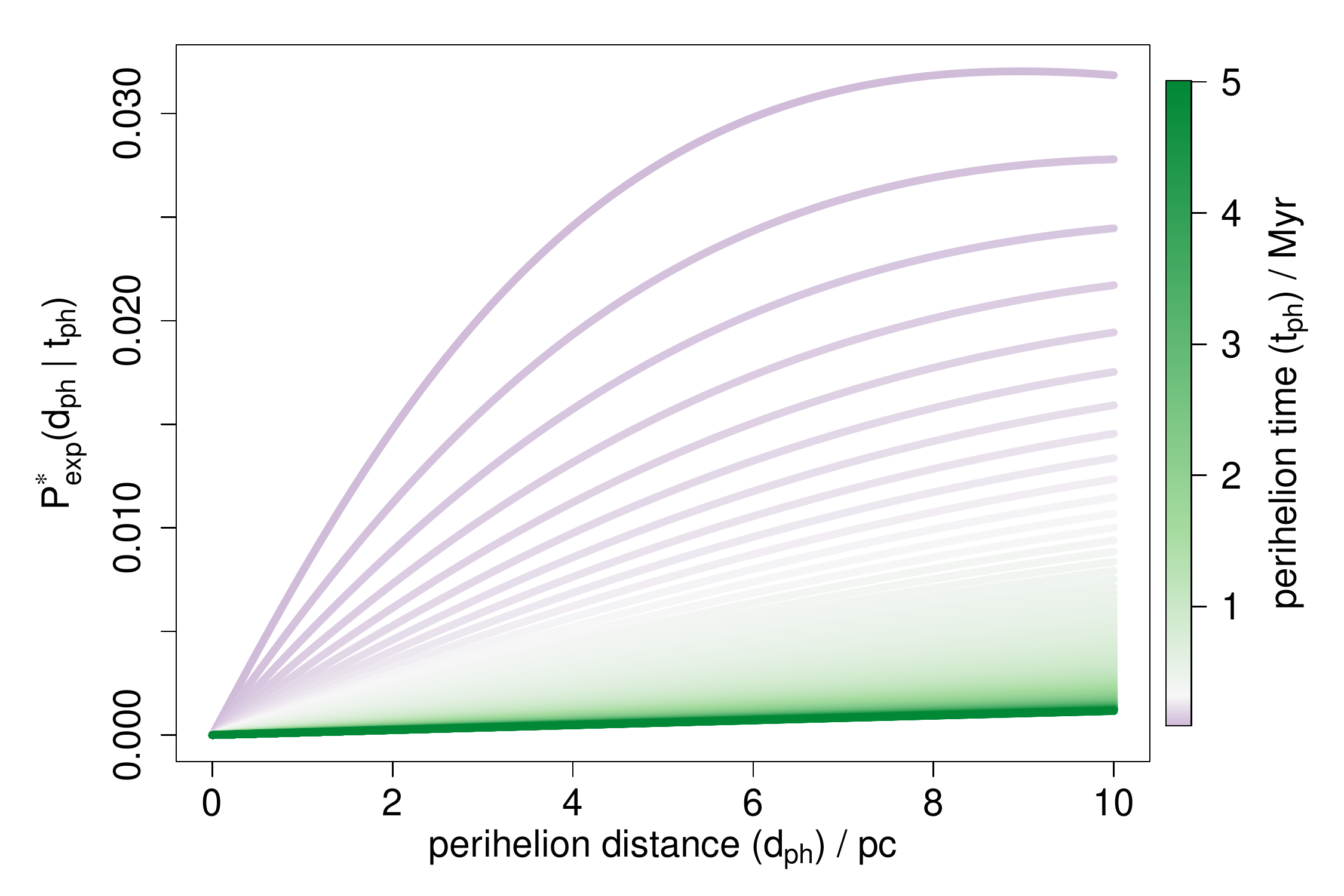}
\includegraphics[width=0.49\textwidth, angle=0]{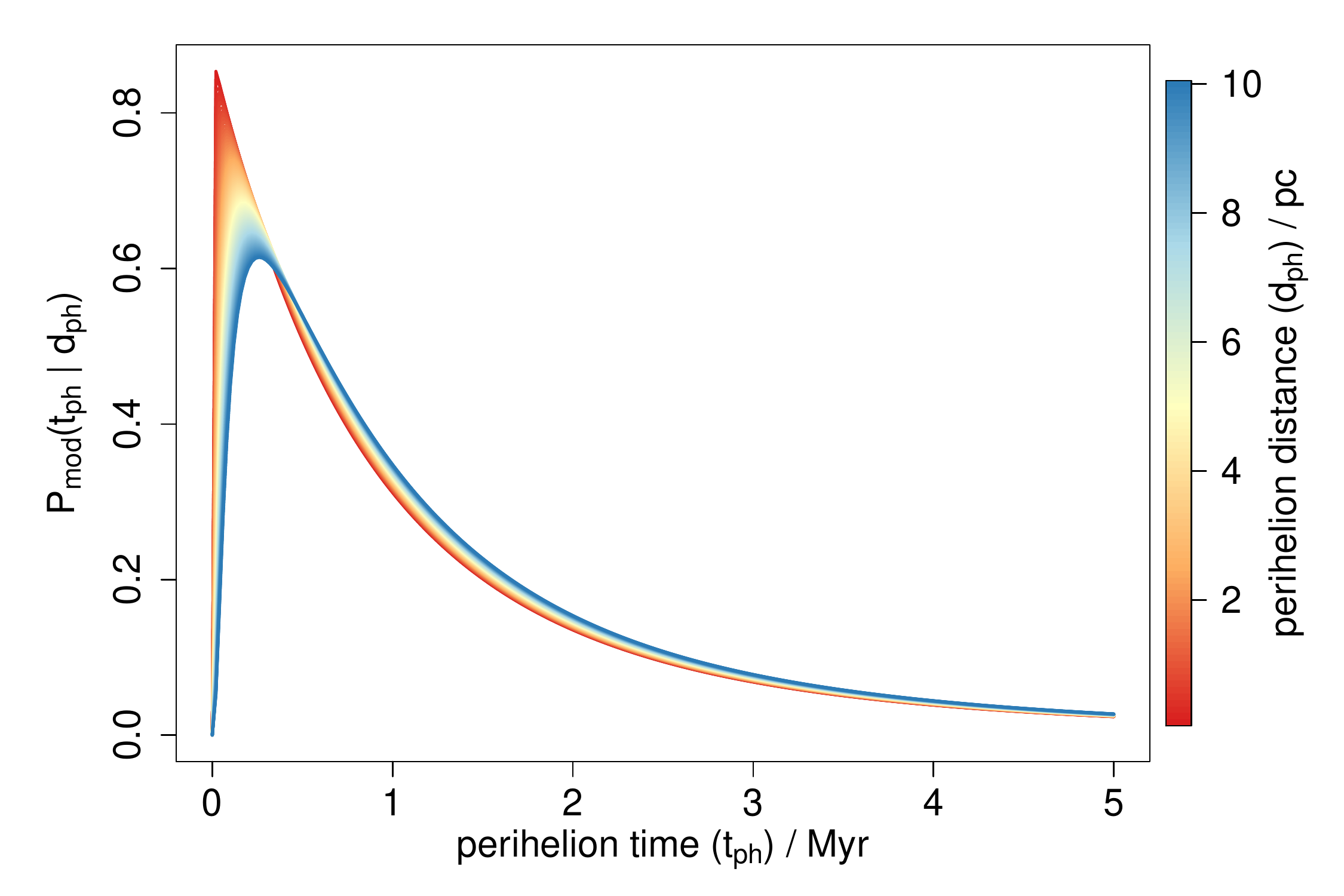}
\includegraphics[width=0.49\textwidth, angle=0]{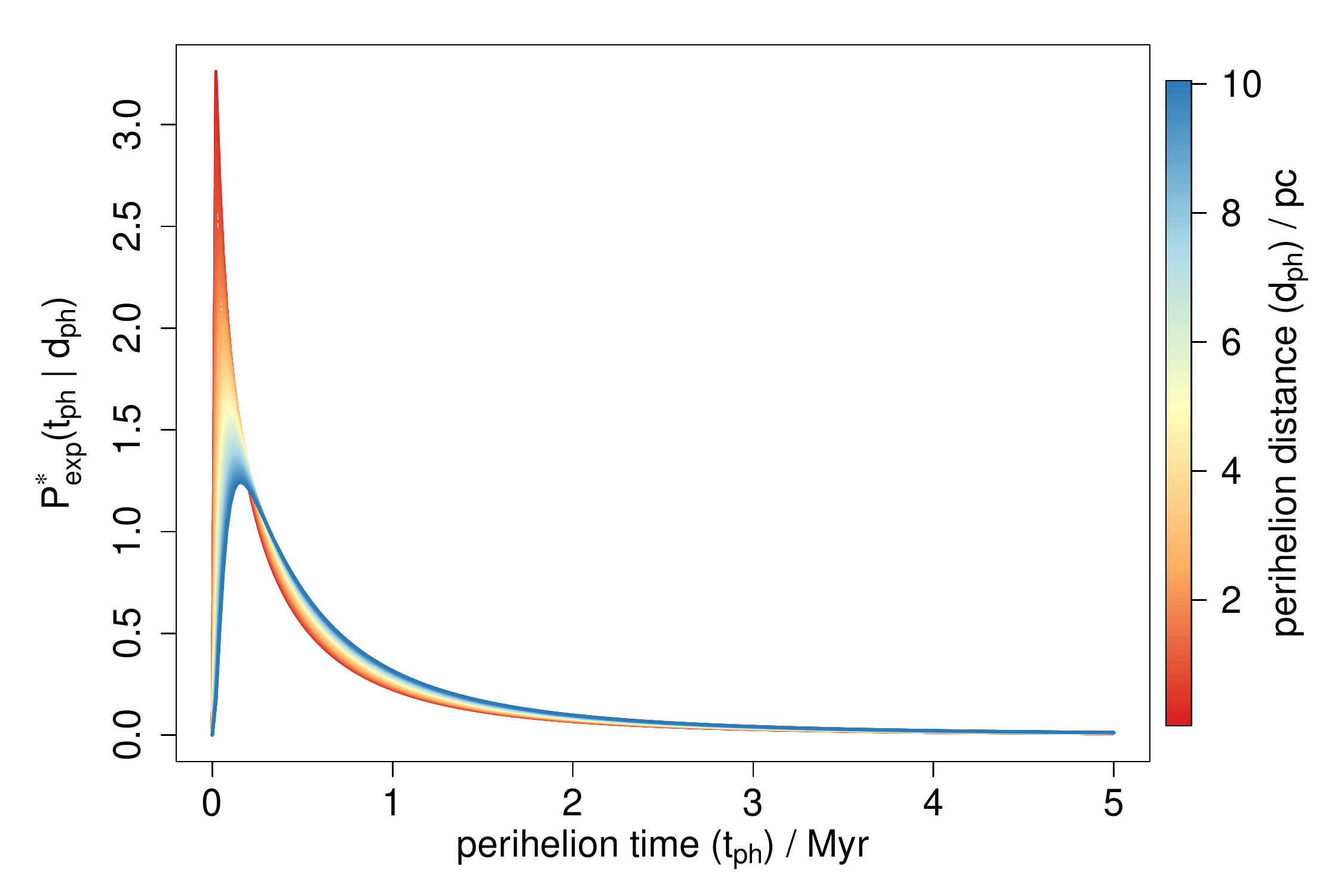}
\caption{The distribution of stellar encounters in the model (left column of plots) and what we expect to observe from this model given the selection function (right column). The top row of plots shows the two-dimensional PDFs $\pmod(\tph, \dph)$ (left) and $\pexp(\tph, \dph)$ (right) on a colour scale in units Myr$^{-1}$pc$^{-1}$. The colour scale covers the full range of densities plotted (and is different in the two cases, being a factor of 4.4 larger for the model).
The middle row shows the one-dimensional conditional PDFs over perihelion distance for different values of $\tph$ ranging from 100\,kyr to 5\,Myr in steps of 20\,kyr. These are vertical cuts through the two-dimensional PDF (and re-normalized, as given by equation \ref{eqn:dph_given_tph} for the model distribution). The bottom row shows the one-dimensional conditional PDFs over perihelion time for different values of $\dph$ ranging from 0.1\,pc to 10\,pc in steps of 0.1\,pc. 
Only the positive perihelion times are shown for the four conditional plots, which also use different density scales (vertical axes).
\label{fig:completeness_model_results}}
\end{center}
\end{figure*}

Using these input distributions, the resulting distribution $\pmod(\tph,\dph)$ for the model is shown in the top left panel of Figure \ref{fig:completeness_model_results}. The distribution we expect to observe -- i.e.\ after application of the selection function -- 
is show in the top right panel. The one-dimensional conditional distributions derived from these, e.g.\
\begin{equation}
\pmod(\dph \given \tph) \,=\, \frac{\pmod(\tph, \dph)}{\int_0^\infty \pmod(\tph, \dph^\prime) \, \deriv\dph^\prime}
\label{eqn:dph_given_tph}
\end{equation}
are shown in the bottom two rows. 
%Note that to compute these, the integral in the denominator must be computed over a much larger grid than that shown in Figure \ref{fig:completeness_model_results}, in order to get an accurate normalization.
%
%\footnote{The denominator for the computation of $\pmod(\tph \given \dph)$ is actually computed instead as $\pmod(\dph) = \int_{\dph}^{\infty} P(\dph \given r)\pmod(r) \, \deriv r$ via Gaussian quadrature.}
% Note: the Pexp conditionals ARE normalized, so strictly I should label these without the asterisk. 

By construction, these distributions are symmetric about $\tph=0$. Consider the form of $\pmod(\tph \given \dph)$ in the bottom left panel of Figure \ref{fig:completeness_model_results}. 
This increases from zero at $\tph=0$ to a maximum somewhere between 0 and 0.25\,Myr (depending on $\dph$) before decreasing at larger $\tph$. 
This can be understood from the encounter geometry and the combination of the input distance and velocity distributions.
Stars that encounter in the very near future are currently nearby (small $r$), but ``nearby'' has a limited volume ($\propto r^3$) and so a limited supply of stars. Thus the further into the future encounters occur, the more volume there is available for their present positions, so -- provided the space density doesn't drop -- the more encounters there can be.
This explains the increase in $\pmod(\tph \given \dph)$ at small $\tph$.
But at large enough distances the space density of stars does drop (Figure \ref{fig:completeness_model_inputs}a),
so the number of potentially encountering stars also decreases. These currently more distant stars generally have encounters further in the future, which is why $\pmod(\tph \given \dph)$ decreases at larger $\tph$.
Stars coming from $\rlen=100$\,pc (where the space density has dropped off by a factor of $e^{-1}$) with the most common space velocity (60\,\kms, the mode of $P(\v)$ in Figure \ref{fig:completeness_model_inputs}b) take 1.7\,Myr to reach us, and we see in the figure that $\pmod(\tph \given \dph)$
has already dropped significantly by such times.\footnote{When using a much larger length scale $\rlen$, and/or a much smaller maximum velocity, then essentially all the encounters that reach us within a few Myr come from regions with the same, constant, space density. 
% completeness/model33
For example, using a velocity distribution with a mode at 20\,\kms, then in 2.5\,Myr stars most stars come from within 50\,pc, out to which distance the space density has dropped by no more than 5\% when using
$\rlen=1000$\,pc.
With this alternative model we still see a rise in $\pmod(\tph \given \dph)$ like that in Figure \ref{fig:completeness_model_results} (now extending up to 1.5\,Myr for the largest $\dph$ on account of the lower speeds). But the curves then remain more or less flat out to 5\,Myr, which reflects the constant space density origin of most of the encountering stars.}

The $P(\tph, \dph)$ distribution is not invariant under translations in the {\em perihelion} time, but this does not mean we are observing at a privileged time.  After all, there is no time evolution of the stellar population in our model, so we would make this same prediction for any time in the past or future.  It is important to realise that Figure \ref{fig:completeness_model_results} shows when and where stars {\em which we observe now} will be at perihelion. 
%(The time axis is not observation time.)
The drop-off at large perihelion times is 
%{\em not} because space is emptied of stars (stars that would then not be available for future generations to observe), but is 
a result of the adopted spatial and velocity distributions.  
In some sense these distributions are not self-consistent, because this velocity distribution would lead to a changing spatial distribution.

Other than for very small values of $\tph$, we see that $\pmod(\dph \given \tph)$ increases linearly with $\dph$ over the perihelion distances shown (0--10\,pc). The number of encounters occurring within a given distance, which is proportional to the integral of this quantity, therefore varies as $\dph^2$.  This behaviour may seem counter-intuitive (e.g.\ for a uniform density, the number of stars within a given distance grows as the cube of the distance).  But it can again be understood when we consider the phase space volume available for encounters: In order for a star to encounter at a distance $\dph$ of the Sun, it must traverse a thin spherical shell (centered on the Sun) of radius $\dph$. The volume of this shell scales as $\dph^2$, so the number of stars traversing this volume also scales as $\dph^2$. Thus the number of encounters per unit perihelion distance, which is proportional to $\pmod(\dph \given \tph)$, scales as $\dph$, as the model predicts.  
% dN = P(dph)ddph  =>  P(dhp) = dN/ddph \propto d/dph(dph^2) = dph
This will break down at large $\dph$, because even though the phase space available for encountering continues to grow, the number of stars available to fill it does not, because of the form of $\pmod(r)$.  Thus $\pmod(\dph \given \tph)$ will decrease for sufficiently large $\dph$ (which anyway must occur, because the PDF is normalized).  Examination of $\pmod(\dph \given \tph)$ out to much larger distances than those plotted confirms this.  There is a second contribution to this turn-over, namely that at smaller $\tph$ the volume from which the encounters can come is also smaller (as explained earlier). Thus for these times 
we begin to ``run out'' of encounters already at smaller values of $\dph$.
% model25
This can be seen as a levelling-off of $\pmod(\dph \given \tph)$ at very small $\tph$ in Figure \ref{fig:completeness_model_results} (upper lines in middle left panel).
 
The expected distributions (the right column of Figure \ref{fig:completeness_model_results}) are qualitatively similar to the model distributions. The main difference is that the expected distribution is squeezed toward closer times. This is consistent with the fact that we only observe the brighter and therefore generally nearer stars in our survey, which correspondingly have smaller encounter times. 

\begin{figure}
\begin{center}
\includegraphics[width=0.5\textwidth, angle=0]{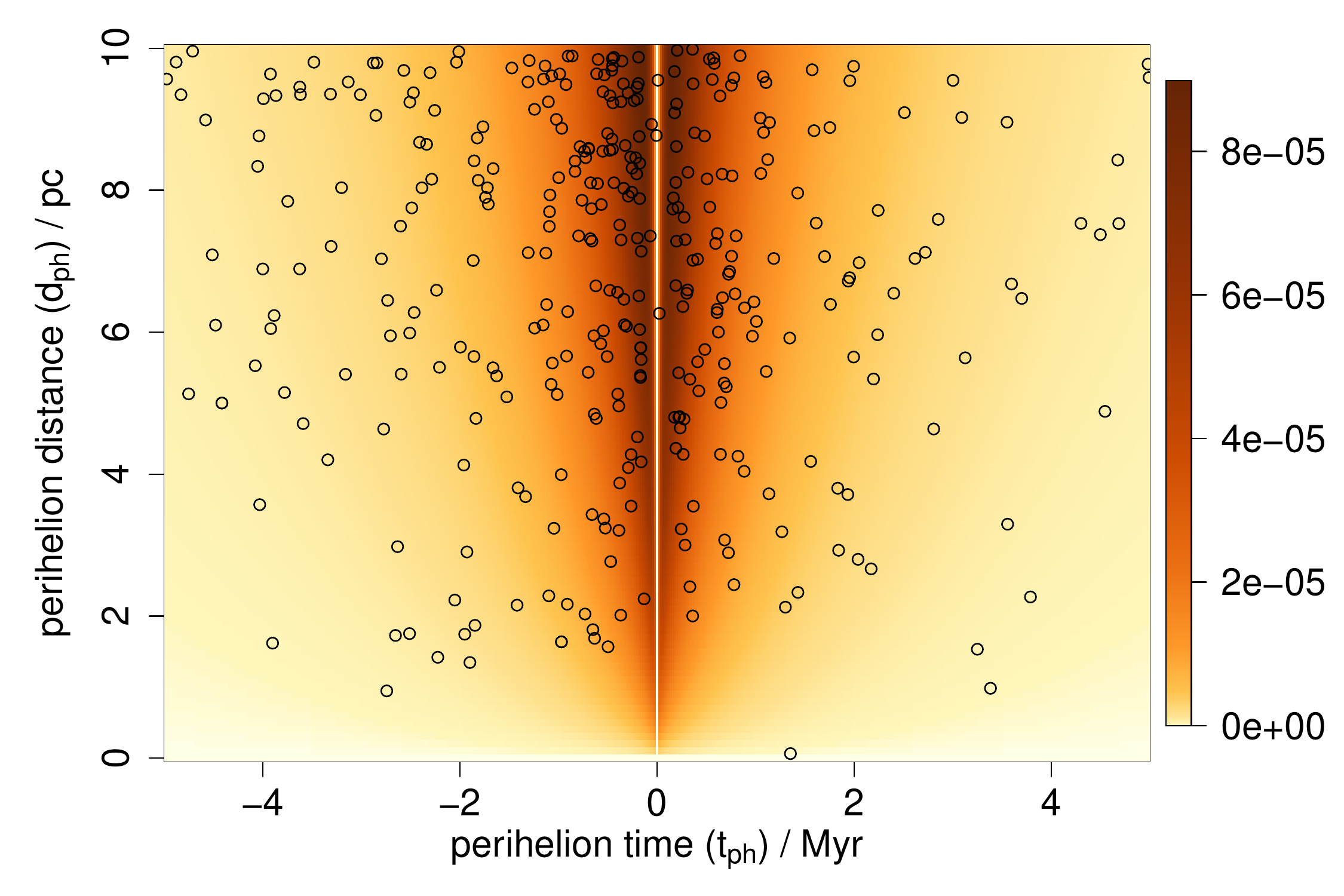}
\caption{As the top right panel of Figure \ref{fig:completeness_model_results}, now overplotted with the TGAS encounters found using the linear motion approximation $(\tphlin, \dphlin)$, with duplicates removed at random.
\label{fig:completeness_with_LMA_encounters}}
\end{center}
\end{figure}

Figure \ref{fig:completeness_with_LMA_encounters} overplots the perihelion times and distances of the TGAS encounters with this expected distribution.  For this plot I have used the encounters as found with the linear motion approximation, rather than with the orbit integrations, because the latter will be incomplete near to 10\,pc (as explained in section \ref{sec:procedure}).  I show only the 439 unique stars (rather than all objects), with duplicates removed at random.
% nrow(enclma)=439 (from completeness/make_nice_plots2.R when using model21 and combined_kincat.Robj)
As the model is very simple we do not expect good agreement with the observations. But the data show a similar distribution, in particular fewer encounters both at large times and at times near to zero.  Recall that the encounter model shows an {\em intrinsic} minimum -- in fact a zero -- at $\tph=0$, i.e.\ it is not due to observational selection effects. Rather it is due to the vanishing amount of phase space available for encountering stars as $\tph$ decreases to zero. So we expect to see this in the data. While we see it here, it is not clearly discernable in my earlier Hipparcos-based study (Figure 2 of paper~1).

\begin{figure}
\begin{center}
\includegraphics[width=0.5\textwidth, angle=0]{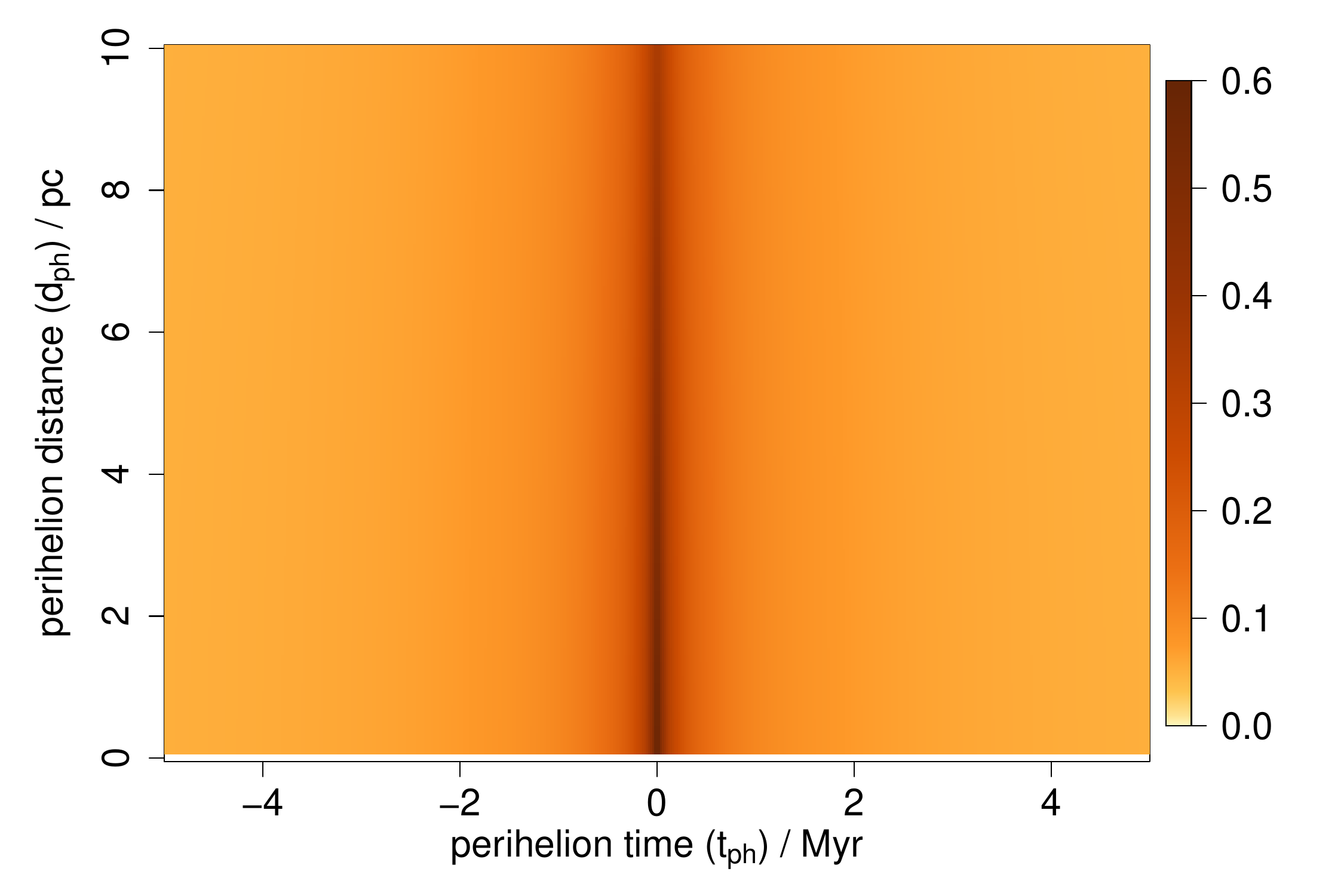}
\includegraphics[width=0.5\textwidth, angle=0]{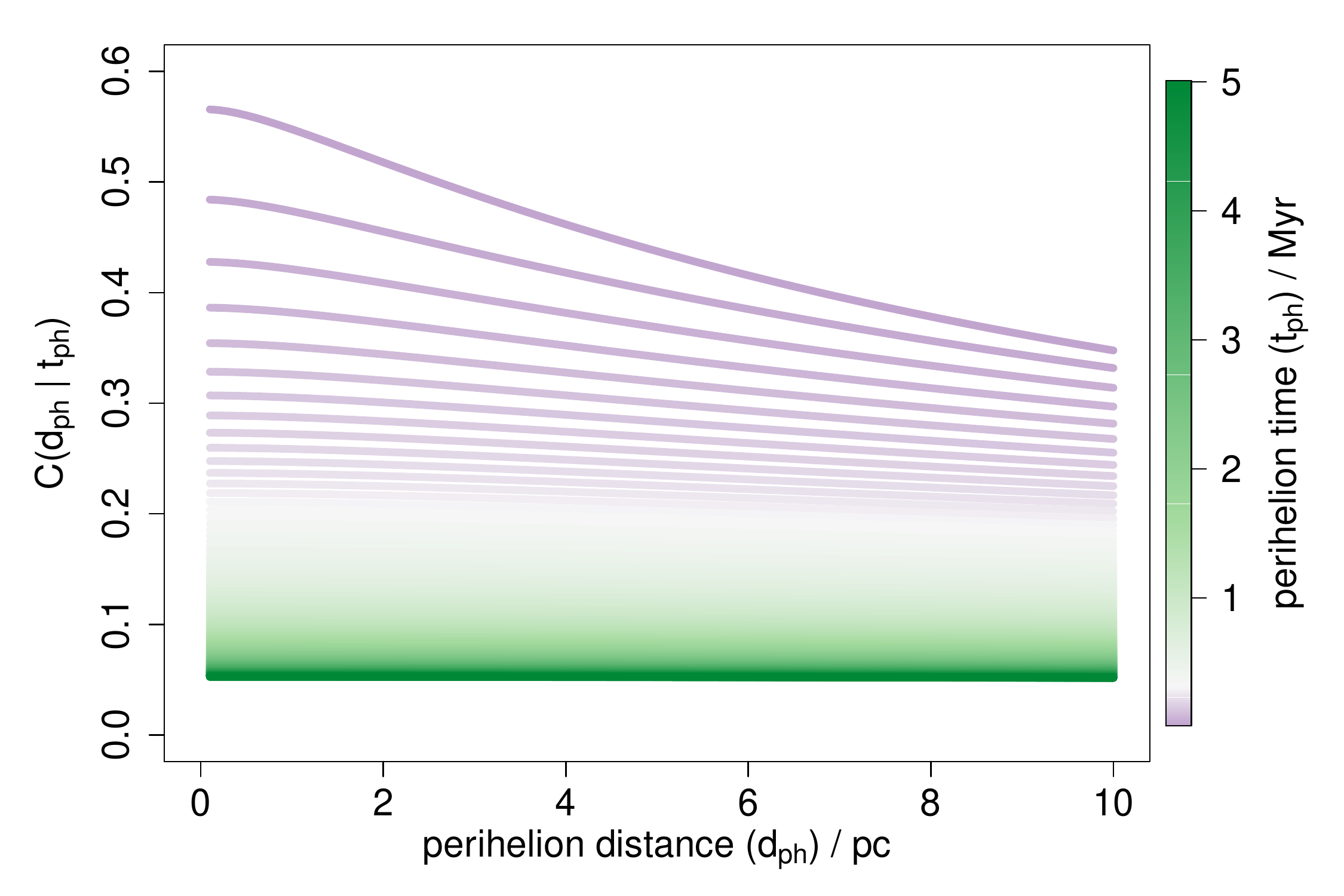}
\includegraphics[width=0.5\textwidth, angle=0]{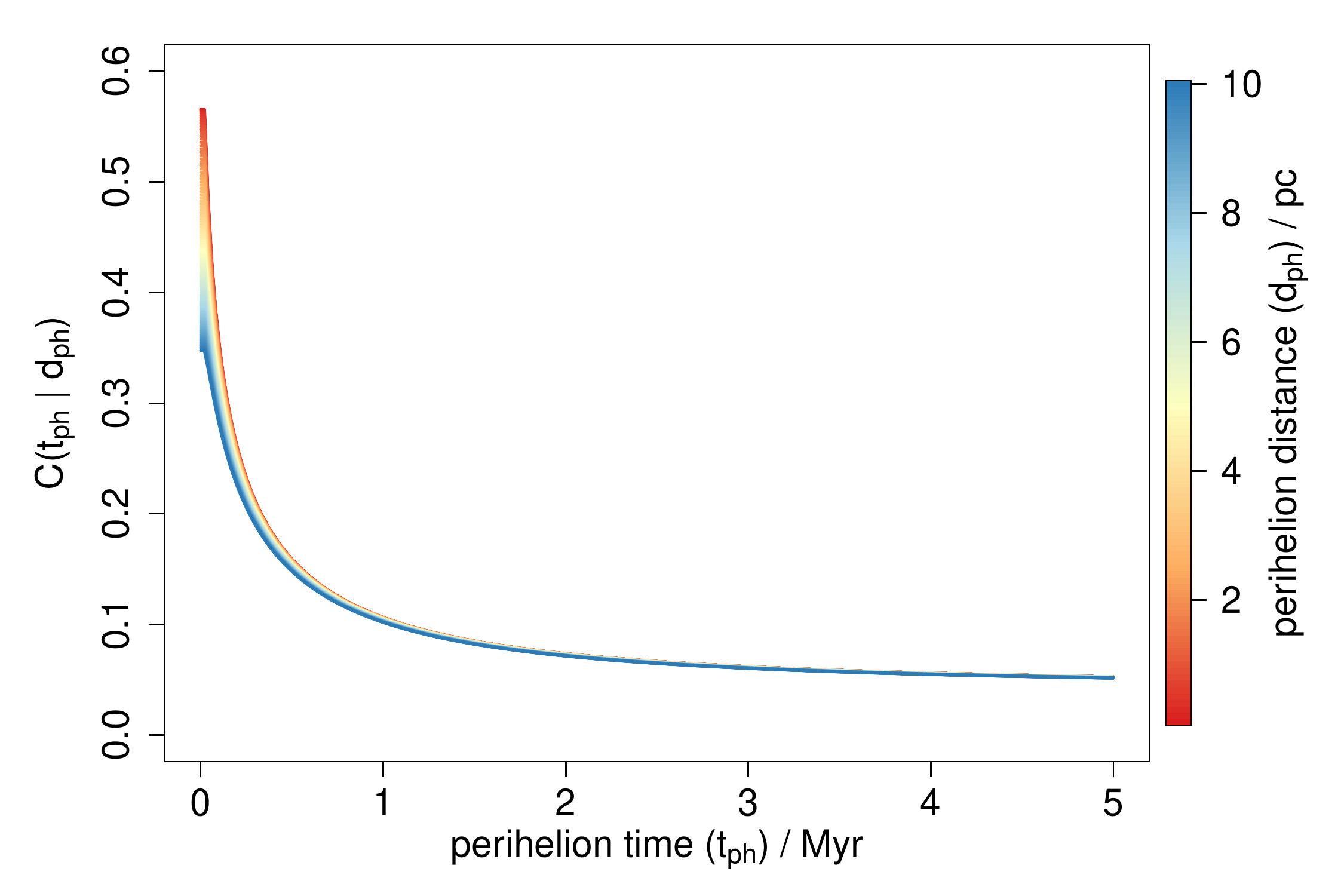}
\caption{The completeness function, $C(\tph, \dph)$ (top panel). This shows the probability of observing an encounter as a function of its perihelion time and distance. It uses distributions of stellar properties and an approximation of the TGAS selection function (see section \ref{sec:themodel}) and is mirror-symmetric about $\tph=0$. The values shown range between 0.05 and 0.57, which excludes $\tph=0$ and $\dph=0$ as formally the incompleteness is undefined at these values.
Cuts through this parallel to the $\dph$ axis are shown in the middle panel for $\tph$ from 20\,kyr to 5\,Myr in steps of 20\,kyr.
Cuts parallel to the $\tph$ axis (for positive $\tph$) are shown in the bottom panel for $\dph$ from 0.1--10\,pc in steps of 0.1\,pc.
\label{fig:C_tph_dph}}
\end{center}
\end{figure}

The completeness function (computed with equation \ref{eqn:completeness}) is shown in the top panel of Figure \ref{fig:C_tph_dph}.
%It gives the probability of observing any particular encounter as a function of the encounter perihelion time and distance. (It does not tell us whether there really is an encounter at that time and distance.) 
%The survey is defined by the selection function $S(m)$ in Figure \ref{fig:completeness_model_inputs}. 
Formally $C(\tph=0, \dph)$ and $C(\tph, \dph=0)$ are undefined (zero divided by zero), but given the time symmetry and the form of the plots, it is clear that at fixed $\dph$ the function varies continuously at $\tph=0$, so I use constant interpolation to fill this.
The true completeness will not go to unity at $\tph=0$ for very small $\dph$, because 
such encountering stars would currently be so close and hence some so bright that they would be unobservable due to the bright limit of the selection function.\footnote{As I don't compute $C$ to arbitrarily small values of $\tph$, the faint magnitude limit also prevents the plotted completeness going to one. For $\tph=20$\,kyr -- the smallest time I use; top line in the middle panel of Figure \ref{fig:C_tph_dph} --  stars moving  at 60\,\kms\ to zero perihelion distance are currently 1.2\,pc away (and faster stars are currently more distant).
Any star at this distance with $M>18.1$ will have $m>13.5$ and so not be observable.}
% Model29 shows that completeness really does not go to 1 at tph=0. 
% Model34 shows it still doesn't go to unity even if we get rid of the bright limit, because in the closest time bin, 20kyr, some stars are still too faint to be currently observable, even at 1.2pc (1.2pc = 60km/s * 20kyr).
% My completeness function does include tph=0 (constant interpolation), but not dph=0
I avoid the formal singularity at $\dph=0$ in the following by using a lower limit of $\dph=0.1$\,pc. 
This and the constant interpolation have negligible impact on the following results.

We see from Figure \ref{fig:C_tph_dph} that the completeness ranges from 0.05 at large perihelion times to 0.57 for very close times.
The mean completeness for $|\tph|<5$\,Myr and $\dph<5$\,pc is 0.091 (0.099 for $\dph<2$\,pc).
% mean(c(compFac[which(tphvec<5e6), which(dphvec<5)], compFac[,2]), na.rm=TRUE)
That is, if all TGAS stars had measured radial velocities, we would expect to observe about 1 in 11 of the encountering stars in this region (if the Galaxy really followed the model assumptions laid out above). 
This might seem a rather large fraction, given that Figure \ref{fig:completeness_model_inputs}(a) shows a much smaller fraction of stars retained when the selection function is applied to $\pmod(r)$ (solid line) to achieve $\pexp(r)$. But $\pexp(r)$ shows the observability of {\em all} stars, not just the encountering stars. A completeness correction based on the observability of all stars would be incorrect, because it would ``overcorrect'' for all those stars which won't be close encounters anyway.

There is very little dependence of the completeness on distance, except at the smallest times, as can be seen in the middle panel of Figure \ref{fig:C_tph_dph}, which shows cuts through the completeness at fixed times.
%(The completeness function is not a density function, so these ``conditional'' plots are not renormalized.)
At $\tph=1$\,Myr, for example, the completeness varies from 0.106 at $\dph=0.1$\,pc to 0.102 at $\dph=10$\,pc. 
% compFac[which(tphvec==1e6)+length(tphvec)-1,which(dphvec==0.1)]
% compFac[which(tphvec==1e6)+length(tphvec)-1,which(dphvec==10)]
There is a strong dependence on time, in contrast, as can be seen in the bottom panel of Figure \ref{fig:C_tph_dph}. 
The lack of distance dependence is because, for a given encounter time, 
those stars which encounter at 0.1\,pc have much the same current spatial distribution as those which encounter at 10\,pc: They all come from a large range of distance which is much larger than 10\,pc (unless $\tph$ is very small), so their average observability and thus completness is much the same.
Perihelion times varying between 0 and 5\,Myr, in contrast, correspond to populations of stars currently at very different distances, for which the observability varies considerably.

\subsection{An estimate of the encounter rate}\label{sec:encounter_frequency}

% The main result is from completeness/model21 using results02and04/sample_multiorbit_combined.Robj, 
% which includes the missing Hipparcos stars

I now use the completeness function to correct the observed distribution of encounters in order to infer the intrinsic (``true'') distribution of encounters. From this we can compute the current encounter rate out to some perihelion distance.

Let $\fint(\tph, \dph)$ be the intrinsic distribution of encounters, i.e.\ the number of encounters per unit perihelion time and perihelion distance with perihelion parameters $(\tph, \dph)$. The corresponding quantity for the observed encounters is 
$\fobs(\tph, \dph)$. These two quantities are the empirical equivalents of $\fmod$ and $\fexp$, respectively, introduced around equation \ref{eqn:cdef1}. This equation therefore tells us that 
\begin{equation}
\fobs(\tph,\dph) \,=\, g C(\tph, \dph) \fint(\tph,\dph)
\label{eqn:fobs}
\end{equation}
where $g$ is an extra factor accommodating the possibility that the search for encounters did not include all the observed stars (discussed below).
Given this factor, the completeness function, and the observed distribution of encounters, we can derive $\fint(\tph,\dph)$.  Integrating this over time and distance we infer the intrinsic number of encounters.

To do this I first represent the observed encounter distribution, $\fobs(\tph, \dph)$, as a continuous density.
I compute this using kernel density estimation (with a Gaussian kernel) on the surrogates for all stars. (The surrogates are the data resamples in the orbital integration -- see item \ref{item:surrogates} in section \ref{sec:procedure}). By applying the density estimation to the surrogates -- rather than, say, the median of the perihelion parameter distribution -- we account for the uncertainties.
%This can and does result in only a fraction of the surrogates for a given star lying within some perihelion region.
When computing the density I only retain unique stars (duplicates are removed at random). I also exclude surrogates with $\vph>300$\,\kms, on the grounds that the
model does not include faster stars either. 
This also gets rid of objects with the most spurious radial velocity measurements.
% If the cut is applied to vphmed then 26 of the 630 unique stars with dphmed<10 are removed.
% If the velocity is raised to 600 km/s, only 5 are cut. (These numbers are prior to adding the extra Hipparcos stars.)

\begin{figure} % completeness/model21
\begin{center}
\includegraphics[width=0.5\textwidth, angle=0]{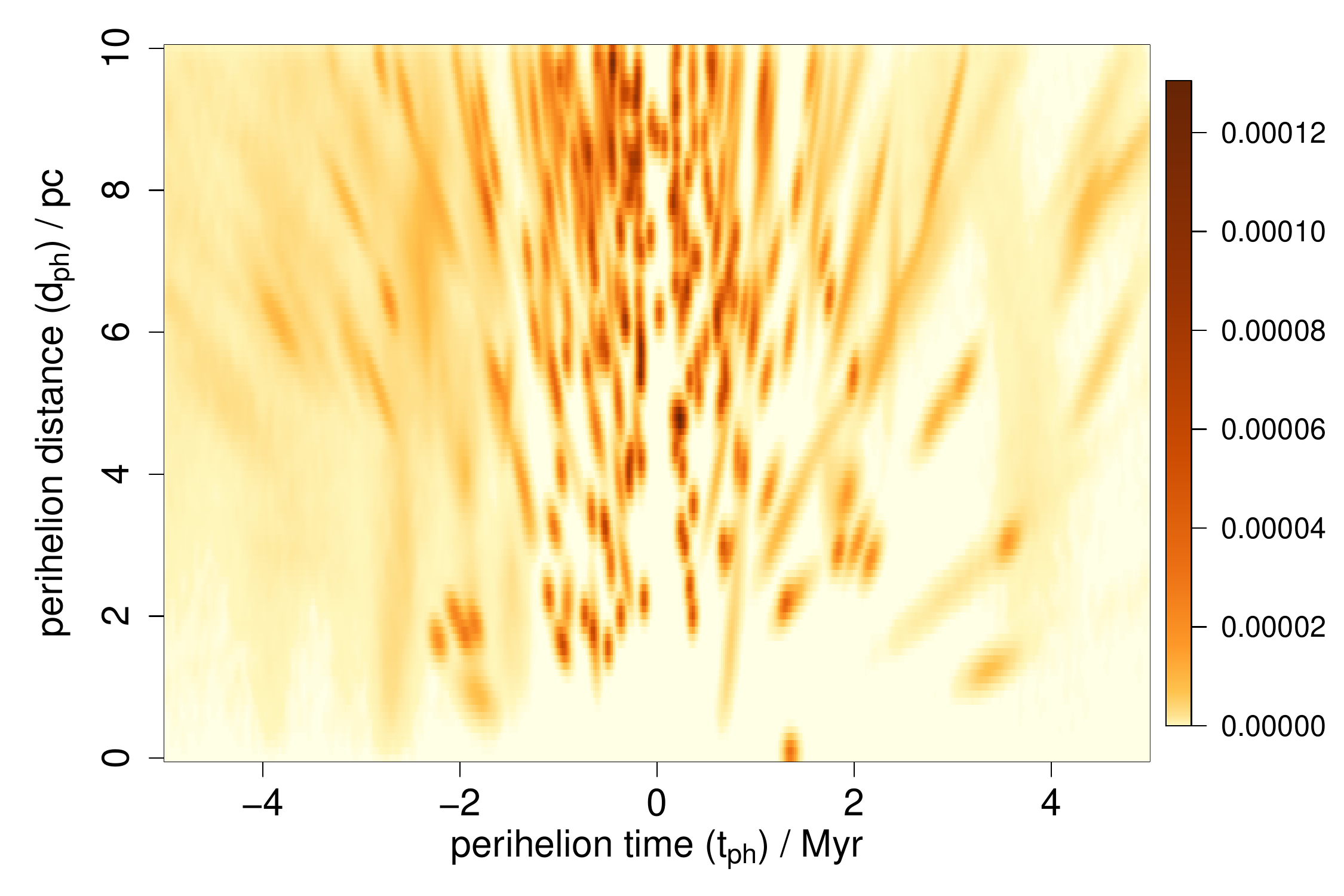}
\caption{The density of the observed encounters, $\fobs(\tph, \dph)$, estimated by kernel density estimation over the surrogates (orbital-integrated data samples) with $\vph \leq 300$\,\kms, for all unique stars. The colour scale gives the number of encounters per unit time and distance. The integral over time and distance is the (fractional) number of encountering stars. 
\label{fig:F_obs_tph_dph}}
\end{center}
\end{figure}

Figure \ref{fig:F_obs_tph_dph} shows the resulting distribution. The density scale has been defined such that the integral over some perihelion time and distance region equals the (fractional) number of stars within that region.  This number can be a fraction because not all of the surrogates for a star lie within the region.
%Remember that the time axis is perihelion time: this plot is {\em not} showing an evolution of the encounter rate with our observation time.

The completeness function describes what fraction of stars we miss on account of the TGAS selection function. But it does not reflect the fact that not all TGAS stars have radial velocities, and so could not even be tested for being potential encounters. This is accommodated by the factor\footnote{Describing this as a constant factor assumes that the selection of radial velocities is not a function of $\tph$ and $\dph$. If fewer faint stars had radial velocities, for example, this {\em may} harm this assumption. We might expect missing faint stars to lower the completeness at larger $\tph$, but it's not as simple as this: the apparently faint stars are at a range of distances, so will contribute to a range of $\tph$. More distant stars are also less likely a priori to encounter the Sun.  Given the ill-defined selection function of the combination of these radial velocity catalogues, any attempt at a more sophisticated correction would probably do more harm than good.}  $g$ in equation \ref{eqn:fobs}.  The total number of stars in TGAS is 2\,057\,050 \citep[][section 5.1]{2016A&A...595A...4L}.
The number of stars in TGAS for which I have valid radial velocities is 322\,462,
and this is approximately the number of stars which I could potentially have searched for encounters.
Therefore, the fraction of stars in TGAS I effectively searched is $g = 322\,462/2\,057\,050 =0.157$. 

I now apply equation \ref{eqn:fobs} to find $\fint(\tph,\dph)$. 
%As I have computed $\fobs(\tph, \dph)$ and $C(\tph, \dph)$ on identical grids, I simply divide them, and then divide the result by $g$. 
Integrating this over some region of $\tph$ and $\dph$ gives the inferred number of encounters in that region. Dividing that by the time range of the region gives the average encounter rate.

For the full region over which the completeness has been computed, $|\tph|\leq 5$\,Myr and $\dph\leq 10$\,pc, the inferred encounter rate is $2117 \pm 115$\,Myr$^{-1}$.  The quoted uncertainty reflects only the Poisson uncertainty arising from the finite number of observed encounters (it is the rate divided by the square root of the fractional number of encounters).
% Origin: the fractional error in the encounter rate should be the same as the fractional error in the number of observed encounters. The latter is sqrt(No)/No = 1/sqrt(No). The derived rate is then multiplied by this.
This rate is likely to be an underestimate, however, because we will miss encounters near to 10\,pc on account of the initial LMA-based selection (explained at the end of section \ref{sec:procedure}).
%and this is an incompleteness this approach does not correct for. 
Limiting the region to perihelion distances within 5\,pc we get $545 \pm 59$\,Myr$^{-1}$. This is 3.9 times smaller, which follows almost exactly the quadratic dependence on the number of encounters within a given distance predicted by the encounter model (see section \ref{sec:completeness_predictions}). 

If I limit the region to 2\,pc we get an encounter rate of $124 \pm 30$\,Myr$^{-1}$. However, this estimate is based primarily on just the 13 stars with $\dphmed<2$\,pc and $\vphmed<300$\,\kms, so has a larger uncertainty. If I instead scale quadratically the result for 5\,pc, I get
% 545*(2/5)^2 
$87 \pm 9$\,Myr$^{-1}$, which is what I adopt as the rate inferred by this study.

The inferred encounter rates for smaller time windows tend to be higher. For $\dph\leq 5$\,pc, the average rates for close encounters within 1, 2, 3, and 4\,Myr of now are $844 \pm 128$, $816 \pm 102$, $725 \pm 83$, and $639 \pm 70$\,Myr$^{-1}$ respectively. These are consistent within the quoted uncertainties, although these estimates are not independent because encounters in the shorter time intervals are also counted in the longer intervals.

\subsection{Other published rates}

Other authors have derived encounter rates using different data sets or from Galaxy models. 
Using the Hipparcos catalogue and a correction for incompleteness, \cite{2001A&A...379..634G} derived a value of
$11.7 \pm 1.3$\,Myr$^{-1}$ within 1\,pc (an average for the past and future 10\,Myr, it seems).
% not clearly stated in paper, but abstract suggests this.
Applying the quadratic scaling with distance, this would correspond to 
$46.8 \pm 5.2$\,Myr$^{-1}$ within 2\,pc, which is about half the size of my result.
Their approach to the incompleteness is very different: they compute a single factor for the fraction of stars within 50\,pc which Hipparcos did not see (estimated as $1/5$), even though many of their encounters would have travelled from much larger distances.
% Any star with v>5km/s would move further than 50pc in the 10Myr time they claim their is derived for. As stars move much faster on average, they are implicitly assuming that the space density of luminosity distribution of stars within 50pc is the same in the much larger volume from which many of the observed encounters must come.
My correction, in contrast, looks at the completeness as a function of encounter time and distance, based on 
non-uniform density and velocity distributions.
The luminosity function they used in their correction (their Figure 13) does not include nearly as many faint stars as mine (Figure \ref{fig:completeness_model_inputs}c). This may explain their smaller inferred encounter rate.

\cite{2017MNRAS.464.2290M} used N-body simulations of the Galaxy
for three different scenarios -- with orbital migration from the inner disk, no significant migration, and with migration from the outer disk -- to derive current encounter rates of 21, 39, and 63\,Myr$^{-1}$ within 1.94\,pc, respectively. These are all somewhat smaller than my inferred rate for the same distance.

\subsection{Sensitivity to assumptions}\label{sec:sensitivity}

My completeness model contains a number of simplifying assumptions, so I look here briefly at their impact on the completeness function and the derived encounter rate.

The adopted distance distribution (equation \ref{eqn:pmod_r}, Figure \ref{fig:completeness_model_inputs}a) uses a length scale of $\rlen=100$\,pc.
% completeness/model22
If I double this, then the model distribution $\pmod(\tph \given \dph)$ shown in the bottom left panel of Figure \ref{fig:completeness_model_results} is stretched in time, because the bulk of the stars are now further away and so take longer to reach perihelion. 
The expected distribution (bottom right panel of Figure \ref{fig:completeness_model_results}) stretches in a similar way. The net result is a decrease in completeness, as we would expect, because the stars are now more distant and so less observable.
The mean completeness for $|\tph|<5$\,Myr and $\dph<5$\,pc is 0.078 (compared to 0.091 before).
Using this completeness function, the inferred encounter rate over the past/future 5\,Myr within 5\,pc is
$659 \pm 71$\,Myr$^{-1}$, compared to 545\,Myr$^{-1}$ computed earlier with $\rlen=100$\,pc. 
% Difference is 1.9sigma, where sigma is uncertainty only on 545 result, as the two measures are entirely correlated.

In the orignal model I assumed that the velocity distribution of stars was as shown in Figure \ref{fig:completeness_model_results}(b). 
% completeness/model23
I now scale this distribution in velocity by a factor of two, keeping the shape the same (i.e.\ the mode shifts from 60\,\kms\ to 120\,\kms). 
% I also change the vphmax threshold from 300 to 600 for computing the encounter rate to be consistent with this
As the stars are now moving much faster, we see relatively more encounters at
closer times in the past/future (and likewise fewer at large times), because their travel times to encounter are shorter.
The conditional distributions are compressed considerably compared to the nominal case.
The mean completeness for $|\tph|<5$\,Myr and $\dph<5$\,pc is now decreased to 0.069, and
the inferred encounter rate over this region is $782 \pm 81$\,Myr$^{-1}$, 1.4 times larger than before.
% completeness/model24 (380 +/- 42)
(Scaling the velocities by a factor of a half -- so the mode is 30\,\kms\ rather than 60\,\kms\ -- lowers the encounter rate by the same factor.)
% and increases the incompleteness to 0.125.

% completeness/model27
In reality, TGAS does not have the sharp bright-magnitude cut I use in the model (Figure \ref{fig:completeness_model_inputs}d). I set this to $G=4.5$ on the grounds that there is only one star brighter in TGAS. Modelling the shape of the completeness at the bright end using very few stars is difficult, so I test the sensitivity to this choice by shifting the cut-off to $G=5.5$ (375 TGAS stars are brighter than this). Remodelling the completeness, I find that the inferred encounter rate increases by just 1\%. This is an increase and not a decrease because while the number of observed encounters stays the same, $\pexp$ and therefore $C$ are slightly smaller, thus increasing $\fint$ (see equations \ref{eqn:completeness} and \ref{eqn:fobs}).
% Query on 2017-06-08:
%  SELECT * FROM gaiadr1.tgas_source AS tgas
%  LEFT OUTER JOIN public.tycho2 AS tycho2
%  ON tgas.hip = tycho2.hip
%  WHERE phot_g_mean_mag < XX
% gives 375 for X=5.5, and 1 for X=4.5 (it is Hip 107418 with phot_g_mean_mag=4.395)

These tests indicate that while the incompleteness-corrected encounter rate does depend on the assumed model parameters, it is not overly sensitive. Of course, the model remains rather unrealistic, not least the assumption of isotropic and independent distance and velocity distributions. Changes to these could have a more dramatic impact. On the other hand, the completeness function depends not on the model directly, but only on how the derived encounter distribution {\em changes} under the observational selection function, which is a weaker dependence. Nonetheless, for the deeper survey we expect in the second Gaia data release, a more sophisticated model should be adopted.

%%%%%%%%%%%%%%%%%%%%%%%%%%%%%%%%%%%%%%%%%%%%%%%%%%%%%%%%%%%%
\section{Summary and conclusions}

I have searched for close stellar encounters using a combination of Gaia astrometry and several radial velocity catalogues.  Candidates were identified from a list of 322\,462 stars 
by assuming that stars travel on straight lines relative to the Sun, and identifying those which come within 10\,pc.  The orbits of these were then integrated in a Galactic potential to compute more precise perihelion parameters.  This included an integration of resampled data (``surrogates'') in order to determine the (asymmetric) distribution over the perihelion time, distance, and speed for each star. These distributions are summarized with the 5\%, 50\% (median), and 95\% percentiles (Table \ref{tab:periStats}).

16 stars were found to have median perihelion distances less than 2\,pc (see Table \ref{tab:dphSummary}).
This is fewer than I found in my Hipparcos-based study (paper~1), and fewer
than expected given the much larger size of TGAS (it has 17 times more stars than Hipparcos, although due to the limited availability of radial velocities it is effectively only four times larger).
This is in part because TGAS includes fewer bright stars (only one brighter than $G=4.5$\,mag), but also because I have excluded stars with large radial velocity uncertainties. Note that some of the encounters listed are almost certainly spurious due, for example, to implausibly large radial velocity measurements.

The closest encounter found is Gl\,710, known for some time to come close to the Sun. TGAS determines a smaller proper motion than Hipparcos for this star, leading to a much smaller perihelion distance: now just 16\,000\,AU (90\% CI 10\,000--21\,000\,AU), making Gl\,710 the closest approaching star known.  This brings it within the Oort cloud, and although its mass is low, so is its velocity, so its perturbing influence is likely to be much higher than more massive but more distant encounters (Figure \ref{fig:dph_vs_tph_impulse}).

I then set up simple models for the stellar spatial, velocity, and luminosity distributions and used these to compute the perihelion distribution we would observe in the absence of observational selection effects (left column of Figure \ref{fig:completeness_model_results}). This shows, for example, a decrease in the number 
of encounters at larger absolute perihelion times. This is a direct (and possibly counter-intuitive) consequence of the adopted spatial and velocity distributions. 
I then determined how this distribution would change under the influence of the TGAS selection function (right column of Figure \ref{fig:completeness_model_results}). 
The ratio of this expected distribution to the model distribution 
gives the completeness function (Figure \ref{fig:C_tph_dph}). This tells us that the probability 
that TGAS observes a star which encounters within 2\,pc within 5\,Myr of now is on average 0.09 (and 0.20 within 1\,Myr from now).
This function shows very little change in the completeness out to 10\,pc perihelion distance, but a strong drop off with perihelion time over $\pm5$\,Myr. This is also seen in the distribution of observed encounters.
I emphasize that the completeness model is based on very simple assumptions, in particular
isotropic spatial and homogeneous velocity distributions, and a selection function which only depends on magnitude. A more complex model would bring few benefits for these data, because the selection functions for TGAS and the RV catalogues would still be poorly defined. The goal in this paper was to explore the consequences of a simple model; this will help the interpretation of more complex models (to be developed for later Gaia data releases). Yet even my simple model is an improvement on earlier attempts at deriving encounter rates, which have either ignored incompleteness or have assumed uniform distributions. Moreover, this is the first study to look at the completeness in the parameter space of interest, which is where we actually need it.

Combining the completeness function with the number of observed encounters (excluding those with $|\vr| > 300$\,\kms), and 
taking into account that not all TGAS stars had radial velocities, I estimated the intrinsic encounter rate.
Averaged over the past/future 5\,Myr this is $545 \pm 59$\,Myr$^{-1}$ for encounters within 5\,pc.
My model predicts that this scales quadratically with the distance limit.
This is confirmed by the data, as the incompleteness-corrected rate out to twice this distance is $2117 \pm 115$\,Myr$^{-1}$, a factor $3.9 \pm 0.4$ larger.
Thus the implied encounter rate within 1\,pc (scaling from the 5\,pc result) is
$21.8 \pm 2.4$\,Myr$^{-1}$. 
The quoted uncertainties reflect only Poisson uncertainties in the detected encounters; uncertainties due to model approximations are probably larger.
The rates given are for all types of stars, as defined by the luminosity distribution in my model (Figure \ref{fig:completeness_model_inputs}c). As the true luminosity distribution is poorly sampled by TGAS, this is one source of additional uncertainty in my derived rates. In particular, TGAS omits more bright stars than my simple selection function accommodates, which would lead to my rates being slightly underestimated.

Some stars which are not both in TGAS and my radial velocity catalogues are additionally known to be close encounters (see references in the introduction). I have not included these in my analysis because cobbling together encounters found in different ways from different surveys makes modelling the incompleteness nearly impossible.
The goal of the present paper was not to list all known close encounters, but rather to make a step toward a more complete modelling with later Gaia data releases. 

%%%%%%%%%%%%%%%%%%%%%%%%%%%%%%%%%%%%%%%%%%%%%%%%%%%%%%%%%%%%
\section{Moving on: future Gaia data releases}\label{sec:future}

The next Gaia data release (GDR2, planned for April 2018) should provide more precise parallaxes and proper motions for of order a billion stars down to $G\simeq21$ and up to $G\simeq4$. This will be vital for increasing the completeness of encounter searches, in particular to encounter times further in the past/future than a few Myr. Using my model together with an estimate of the GDR2 magnitude selection function, 
% model26, which actually used G=3 as a bright limit
I compute the mean completeness over the range $\dph<2$\,pc and $|\tph|<5$\,Myr to be 0.75,
eight times higher than TGAS. The completeness model in the present paper is too simplistic for GDR2, however, because GDR2 will include many more-distant stars. It should also have a better-defined selection function (among other things, it will become independent of Tycho.)
It will then be appropriate to construct a three-dimensional spatial model, as well as a three-dimensional velocity model with spatial dependence, in order to better model the Galaxy (different components, disk rotation, etc.).

The number of encounters which can be found in GDR2 will be limited by the availability of radial velocities. Although GDR2 will include estimates of radial velocities for a few million of the brightest stars (to a precision of a few km/s) -- from the calcium triplet spectra observed by the satellite's high resolution spectrograph \citep{2016A&A...595A...1G} -- this remains a small fraction of stars with astrometry. 
Even upcoming large-scale spectroscopic surveys will obtain ``only'' tens of millions of spectra.
We are now entering a period in which the ability to find encounters is limited not by the availability of astrometry, but by the availability of radial velocities. Furthermore, the Gaia astrometric accuracy is so good that the corresponding transverse velocity accuracy for encounter candidates will far exceed the precision of large-scale radial velocity surveys. Using the official predictions of the end-of-5-year-mission Gaia accuracy\footnote{\url{https://www.cosmos.esa.int/web/gaia/science-performance}}, a G-type dwarf at a distance of 200\,pc moving with a transverse velocity of 5\,\kms\ will have this velocity determined to an accuracy of 8\,\ms.
% tgas_paper2/vt_precision.R

The following (third) Gaia data release will also include estimates of stellar masses and radii, derived by the DPAC using the low resolution Gaia spectroscopy, magnitude, parallax, and stellar models \citep{2013A&A...559A..74B}. We will then be able to estimate systematically actual impulses (equation \ref{eqn:impulse}), and also to correct the measured radial velocities for the gravitational redshift, which is of order 0.5--1.0\,\kms\ \citep{2011A&A...526A.127P}. This data release should also include solutions for astrometric and spectroscopic binaries \citep{2011AIPC.1346..122P}, thereby permitting a better computation of flight paths for such systems.

%%%%%%%%%%%%%%%%%%%%%%%%%%%%%%%%%%%%%%%%%%%%%%%%%%%%%%%%%%%%
\begin{acknowledgements}

I thank members of the MPIA Gaia group for useful comments, and in particular Morgan Fouesneau for discussions on the completeness function and for assistance with ADQL. I also thank the anonymous referee for a constructively critical reading.
This work is based on data from the European Space Agency (ESA) mission {\it Gaia} (\url{https://www.cosmos.esa.int/gaia}), processed by the {\it Gaia} Data Processing and Analysis Consortium (DPAC, \url{https://www.cosmos.esa.int/web/gaia/dpac/consortium}). Funding for the DPAC has been provided by national institutions, in particular the institutions participating in the {\it Gaia} Multilateral Agreement.
This work has used the VizieR catalogue access tool, the Simbad object database, and the cross-match service, all provided by CDS, Strasbourg, as well as TOPCAT.
Funding for RAVE has been provided by: the Australian Astronomical Observatory; the Leibniz-Institut fuer Astrophysik Potsdam (AIP); the Australian National University; the Australian Research Council; the French National Research Agency; the German Research Foundation (SPP 1177 and SFB 881); the European Research Council (ERC-StG 240271 Galactica); the Istituto Nazionale di Astrofisica at Padova; The Johns Hopkins University; the National Science Foundation of the USA (AST-0908326); the W.M.\ Keck foundation; the Macquarie University; the Netherlands Research School for Astronomy; the Natural Sciences and Engineering Research Council of Canada; the Slovenian Research Agency; the Swiss National Science Foundation; the Science \& Technology Facilities Council of the UK; Opticon; Strasbourg Observatory; and the Universities of Groningen, Heidelberg, and Sydney.  The RAVE web site is \url{https://www.rave-survey.org}.

\end{acknowledgements}

\bibliographystyle{aa}
\bibliography{stellar_encounters_tgas}

\appendix

\section{Gaia archive query}\label{appendix:query}

Below is the ADQL query used to select stars from the TGAS table which have perihelion distances less than 10\,pc according to the linear motion approximation, assuming they all have $|\vr|=750$\,\kms.
The TGAS table does not list Tycho IDs when Hipparcos IDs are present,
so I use a join to add these missing Tycho IDs.
\small{
\begin{verbatim}
select tgas.tycho2_id, tycho2.id, tgas.hip, 
tgas.source_id, tgas.phot_g_mean_mag, tgas.ra, 
tgas.dec, tgas.parallax, tgas.pmra, tgas.pmdec, 
tgas.ra_error, tgas.dec_error, tgas.parallax_error, 
tgas.pmra_error, tgas.pmdec_error, tgas.ra_dec_corr, 
tgas.ra_parallax_corr, tgas.ra_pmra_corr, 
tgas.ra_pmdec_corr, tgas.dec_parallax_corr, 
tgas.dec_pmra_corr, tgas.dec_pmdec_corr, 
tgas.parallax_pmra_corr, tgas.parallax_pmdec_corr, 
tgas.pmra_pmdec_corr
from gaiadr1.tgas_source as tgas
left outer join public.tycho2 as tycho2
on tgas.hip = tycho2.hip
where( ( 1000*4.74047*sqrt(power(tgas.pmra,2) + 
  power(tgas.pmdec,2))/power(tgas.parallax,2) ) /
  ( sqrt( (power(tgas.pmra,2) + power(tgas.pmdec,2)) * 
  power(4.74047/tgas.parallax,2) + power(750,2) ) )
  ) < 10
\end{verbatim}
}

\section{The completeness function}\label{appendix:completeness}

Let $\fmod(\tph,\dph)$ be the model-predicted number of encounters per unit perihelion time and perihelion distance
at $(\tph, \dph)$, and let
$\fexp(\tph,\dph)$ be the corresponding expected quantity, i.e.\ after modulation by the observational selection function.
The completeness function, $C(\tph, \dph)$, specifies the fraction of objects observed
at a given $(\tph, \dph)$, and is defined by
\begin{equation}
\fexp(\tph,\dph) \,=\, C(\tph, \dph) \fmod(\tph,\dph) \ .
\label{eqn:cdef2}
\end{equation}
Clearly $0 \leq C(\tph, \dph) \leq 1$.
$\pmod(\tph,\dph)$ is the normalized PDF of $\fmod(\tph,\dph)$, i.e.\
\begin{equation}
\pmod(\tph, \dph) \,=\, \frac{\fmod(\tph, \dph)}{\iint \fmod(\tph^\prime, \dph^\prime) \,\deriv\tph^\prime\deriv\dph^\prime} \,=\,  \frac{\fmod(\tph, \dph)}{\nmod}
\label{eqn:pmod_tphdph}
\end{equation}
where the integral is over all perihelion times and distances, which (conceptually) gives the total number of encounters in the model, $\nmod$. (This is only a conceptual definition for $\nmod$, because my model is continuous.) $\pexp(\tph,\dph)$ relates to $\fexp(\tph,\dph)$ in a similar way, but here we must include an additional (dimensionless) factor, $b$, to accommodate the fact that $\pexp(\tph,\dph)$ is not normalized, 
\begin{equation}
\pexp(\tph, \dph) \,=\, \frac{b \, \fexp(\tph, \dph)}{\iint \fexp(\tph^\prime, \dph^\prime) \,\deriv\tph^\prime\deriv\dph^\prime} \,=\,  \frac{b \, \fexp(\tph, \dph)}{\nexp} 
\label{eqn:pexp_tphdph}
\end{equation}
which also defines $\nexp$.
Integrating both sides of this equation over all perihelion times and distances shows us that
\begin{equation}
b \,=\, \iint \pexp(\tph^\prime, \dph^\prime) \, \deriv\tph^\prime\deriv\dph^\prime \ .
\label{eqn:beval}
\end{equation}
By construction $\pexp(\tph, \dph)$ is the reduction in $\pmod(\tph, \dph)$ under the influence of the selection function, so 
$\pexp(\tph, \dph) \leq \pmod(\tph, \dph)$ everywhere. Hence $b$, which therefore lies between 0 and 1, 
is the fraction of stars retained following application of the selection function, which is $\nexp/\nmod$.

Substituting $\fmod(\tph, \dph)$ from equation \ref{eqn:pmod_tphdph} and $\fexp(\tph, \dph)$ from \ref{eqn:pexp_tphdph} into equation \ref{eqn:cdef2} gives
\begin{equation}
C(\tph, \dph) \,=\, \frac{\pexp(\tph, \dph)}{\pmod(\tph,\dph)}\frac{\nexp}{\nmod}\frac{1}{b} \ .
\label{eqn:cres1}
\end{equation}
But $b=\nexp/\nmod$, so
\begin{equation}
C(\tph, \dph) \,=\, \frac{\pexp(\tph, \dph)}{\pmod(\tph,\dph)} \ .
\label{eqn:cres2}
\end{equation}

\end{document}

%% file: figures/encounters_2pc.tex
% Tycho & cat & tph/kyr & 5% & 95% & dph/pc & 5% & 95% & vph/km/s & 5% & 95% & <0.5pc & <1pc & <2pc & par/mas & epar/mas & pm/mas/yr & epm/mas/yr & rv/km/s & erv/km/s \myeol 
  5102-100-1 &  3 &     1354 &     1304 &     1408 &   0.08 &   0.05 &   0.10 &    13.8 &    13.3 &    14.3 & 100 & 100 & 100 &    52.35 &     0.27 &     0.50 &     0.17 &   -13.8 &     0.3 \myeol 
 4744-1394-1 &  1 &    -1821 &    -2018 &    -1655 &   0.87 &   0.61 &   1.20 &   121.0 &   118.6 &   123.6 &   1 &  74 & 100 &     4.44 &     0.26 &     0.59 &     0.12 &   120.7 &     1.6 \myeol 
  1041-996-1 &  3 &     3386 &     3163 &     3644 &   1.26 &   1.07 &   1.50 &    30.4 &    29.9 &    30.9 &   0 &   1 & 100 &     9.51 &     0.41 &     0.57 &     0.05 &   -30.4 &     0.3 \myeol 
  5033-879-1 &  1 &     -704 &     -837 &     -612 &   1.28 &   0.48 &   2.63 &   532.7 &   522.7 &   543.2 &   6 &  30 &  83 &     2.60 &     0.25 &     0.91 &     0.78 &   532.4 &     6.2 \myeol 
    709-63-1 &  3 &     -497 &     -505 &     -490 &   1.56 &   1.53 &   1.59 &    22.2 &    21.9 &    22.5 &   0 &   0 & 100 &    87.66 &     0.29 &    56.44 &     0.09 &    22.0 &     0.2 \myeol 
 4753-1892-2 &  3 &     -948 &     -988 &     -910 &   1.58 &   1.46 &   1.72 &    55.0 &    54.5 &    55.5 &   0 &   0 & 100 &    18.75 &     0.47 &     6.52 &     0.05 &    55.0 &     0.3 \myeol 
 4753-1892-1 &  3 &     -949 &     -990 &     -908 &   1.59 &   1.46 &   1.73 &    55.0 &    54.5 &    55.5 &   0 &   0 & 100 &    18.75 &     0.47 &     6.52 &     0.05 &    55.0 &     0.3 \myeol 
 4753-1892-2 &  2 &     -970 &    -1027 &     -921 &   1.62 &   1.49 &   1.78 &    53.7 &    51.6 &    55.9 &   0 &   0 &  99 &    18.75 &     0.47 &     6.52 &     0.05 &    53.7 &     1.3 \myeol 
 4753-1892-1 &  2 &     -971 &    -1028 &     -915 &   1.63 &   1.48 &   1.78 &    53.8 &    51.5 &    55.9 &   0 &   0 & 100 &    18.75 &     0.47 &     6.52 &     0.05 &    53.7 &     1.3 \myeol 
  4855-266-1 &  3 &    -2222 &    -2297 &    -2153 &   1.67 &   1.51 &   1.84 &    34.6 &    34.1 &    35.1 &   0 &   0 &  99 &    12.74 &     0.24 &     1.68 &     0.13 &    34.5 &     0.3 \myeol 
    8560-8-1 &  1 &     -634 &     -656 &     -614 &   1.72 &   1.25 &   2.26 &    86.5 &    85.4 &    87.4 &   0 &   0 &  80 &    17.84 &     0.33 &     9.79 &     1.78 &    86.4 &     0.6 \myeol 
 6510-1219-1 &  3 &    -1951 &    -2021 &    -1880 &   1.74 &   1.66 &   1.81 &    14.6 &    14.2 &    15.1 &   0 &   0 & 100 &    34.19 &     0.29 &     6.29 &     0.04 &    14.6 &     0.3 \myeol 
  5383-187-1 & 10 &     -870 &    -1324 &     -651 &   1.74 &   0.40 &   6.23 &   686.8 &   683.2 &   690.1 &   7 &  24 &  56 &     1.63 &     0.36 &     0.55 &     0.57 &   686.7 &     2.1 \myeol 
 6510-1219-1 &  2 &    -1975 &    -2028 &    -1924 &   1.76 &   1.69 &   1.82 &    14.5 &    14.1 &    14.8 &   0 &   0 & 100 &    34.19 &     0.29 &     6.29 &     0.04 &    14.4 &     0.2 \myeol 
 1315-1871-1 &  3 &     -651 &     -663 &     -640 &   1.78 &   1.72 &   1.84 &    40.6 &    40.4 &    40.7 &   0 &   0 & 100 &    36.93 &     0.40 &    21.13 &     0.04 &    40.5 &     0.1 \myeol 
 1315-1871-1 &  2 &     -651 &     -663 &     -640 &   1.78 &   1.72 &   1.84 &    40.6 &    40.4 &    40.7 &   0 &   0 & 100 &    36.93 &     0.40 &    21.13 &     0.04 &    40.5 &     0.1 \myeol 
  6975-656-1 &  1 &     -457 &     -554 &     -390 &   1.80 &   0.89 &   3.89 &   563.1 &   547.2 &   578.3 &   0 &   9 &  58 &     3.79 &     0.39 &     2.76 &     1.72 &   562.9 &     9.2 \myeol 
 4771-1201-1 &  3 &    -1848 &    -1917 &    -1783 &   1.89 &   1.70 &   2.09 &    22.2 &    21.6 &    22.9 &   0 &   0 &  82 &    23.79 &     0.33 &     4.96 &     0.03 &    22.2 &     0.4 \myeol 
  9327-264-1 &  1 &    -1889 &    -2014 &    -1777 &   1.91 &   1.10 &   3.18 &    52.7 &    51.3 &    54.2 &   0 &   3 &  56 &     9.83 &     0.34 &     1.44 &     0.82 &    52.4 &     0.9 \myeol 
  7068-802-1 &  1 &    -2641 &    -2857 &    -2455 &   1.99 &   0.54 &   4.59 &    65.5 &    62.6 &    68.4 &   4 &  17 &  50 &     5.67 &     0.23 &     0.76 &     0.66 &    65.0 &     1.8 \myeol 

%% file: stellar_encounters_tgas.bbl
\begin{thebibliography}{47}
\expandafter\ifx\csname natexlab\endcsname\relax\def\natexlab#1{#1}\fi

\bibitem[{{Bailer-Jones}(2015{\natexlab{a}})}]{2015A&A...575A..35B}
{Bailer-Jones}, C.~A.~L. 2015{\natexlab{a}}, \aap, 575, A35

\bibitem[{{Bailer-Jones}(2015{\natexlab{b}})}]{2015PASP..127..994B}
{Bailer-Jones}, C.~A.~L. 2015{\natexlab{b}}, \pasp, 127, 994

\bibitem[{{Bailer-Jones}(2017)}]{2017IAUS..331..xxxB}
{Bailer-Jones}, C.~A.~L. 2017, in IAU Symposium, Vol. 330, Astronomy and
  astrophysics in the Gaia sky, ed. A.~{Recio-Blanco}, P.~{de Laverny},
  A.~{Brown}, \& T.~{Prusti}, xxx

\bibitem[{{Bailer-Jones} {et~al.}(2013){Bailer-Jones}, {Andrae}, {Arcay},
  {Astraatmadja}, {Bellas-Velidis}, {Berihuete}, {Bijaoui}, {Carri{\'o}n},
  {Dafonte}, {Damerdji}, {Dapergolas}, {de Laverny}, {Delchambre}, {Drazinos},
  {Drimmel}, {Fr{\'e}mat}, {Fustes}, {Garc{\'{\i}}a-Torres}, {Gu{\'e}d{\'e}},
  {Heiter}, {Janotto}, {Karampelas}, {Kim}, {Knude}, {Kolka}, {Kontizas},
  {Kontizas}, {Korn}, {Lanzafame}, {Lebreton}, {Lindstr{\o}m}, {Liu},
  {Livanou}, {Lobel}, {Manteiga}, {Martayan}, {Ordenovic}, {Pichon},
  {Recio-Blanco}, {Rocca-Volmerange}, {Sarro}, {Smith}, {Sordo}, {Soubiran},
  {Surdej}, {Th{\'e}venin}, {Tsalmantza}, {Vallenari}, \&
  {Zorec}}]{2013A&A...559A..74B}
{Bailer-Jones}, C.~A.~L., {Andrae}, R., {Arcay}, B., {et~al.} 2013, \aap, 559,
  A74

\bibitem[{{Barbier-Brossat} \& {Figon}(2000)}]{2000A&AS..142..217B}
{Barbier-Brossat}, M. \& {Figon}, P. 2000, \aaps, 142, 217

\bibitem[{{Berski} \& {Dybczy{\'n}ski}(2016)}]{2016A&A...595L..10B}
{Berski}, F. \& {Dybczy{\'n}ski}, P.~A. 2016, \aap, 595, L10

\bibitem[{{Bobylev}(2010{\natexlab{a}})}]{2010AstL...36..220B}
{Bobylev}, V.~V. 2010{\natexlab{a}}, Astronomy Letters, 36, 220

\bibitem[{{Bobylev}(2010{\natexlab{b}})}]{2010AstL...36..816B}
{Bobylev}, V.~V. 2010{\natexlab{b}}, Astronomy Letters, 36, 816

\bibitem[{{Bobylev} \& {Bajkova}(2017)}]{2017arXiv170608867B}
{Bobylev}, V.~V. \& {Bajkova}, A.~T. 2017, ArXiv e-prints 1706.08867

\bibitem[{{Casagrande} {et~al.}(2011){Casagrande}, {Sch{\"o}nrich}, {Asplund},
  {Cassisi}, {Ram{\'{\i}}rez}, {Mel{\'e}ndez}, {Bensby}, \&
  {Feltzing}}]{2011A&A...530A.138C}
{Casagrande}, L., {Sch{\"o}nrich}, R., {Asplund}, M., {et~al.} 2011, \aap, 530,
  A138

\bibitem[{{Duflot} {et~al.}(1995){Duflot}, {Figon}, \&
  {Meyssonnier}}]{1995A&AS..114..269D}
{Duflot}, M., {Figon}, P., \& {Meyssonnier}, N. 1995, \aaps, 114, 269

\bibitem[{{Dybczynski}(1994)}]{1994CeMDA..58..139D}
{Dybczynski}, P.~A. 1994, Celestial Mechanics and Dynamical Astronomy, 58, 139

\bibitem[{{Dybczy{\'n}ski}(2002)}]{2002A&A...396..283D}
{Dybczy{\'n}ski}, P.~A. 2002, \aap, 396, 283

\bibitem[{{Dybczy{\'n}ski}(2006)}]{2006A&A...449.1233D}
{Dybczy{\'n}ski}, P.~A. 2006, \aap, 449, 1233

\bibitem[{{Dybczy{\'n}ski} \& {Berski}(2015)}]{2015MNRAS.449.2459D}
{Dybczy{\'n}ski}, P.~A. \& {Berski}, F. 2015, \mnras, 449, 2459

\bibitem[{{Famaey} {et~al.}(2005){Famaey}, {Jorissen}, {Luri}, {Mayor}, {Udry},
  {Dejonghe}, \& {Turon}}]{2005A&A...430..165F}
{Famaey}, B., {Jorissen}, A., {Luri}, X., {et~al.} 2005, \aap, 430, 165

\bibitem[{{Fehrenbach} {et~al.}(1997){Fehrenbach}, {Duflot}, {Mannone},
  {Burnage}, \& {Genty}}]{1997A&AS..124..255F}
{Fehrenbach}, C., {Duflot}, M., {Mannone}, C., {Burnage}, R., \& {Genty}, V.
  1997, \aaps, 124

\bibitem[{{Feng} \& {Bailer-Jones}(2014)}]{2014MNRAS.442.3653F}
{Feng}, F. \& {Bailer-Jones}, C.~A.~L. 2014, \mnras, 442, 3653

\bibitem[{{Feng} \& {Bailer-Jones}(2015)}]{2015MNRAS.454.3267F}
{Feng}, F. \& {Bailer-Jones}, C.~A.~L. 2015, \mnras, 454, 3267

\bibitem[{{Fouchard} {et~al.}(2011){Fouchard}, {Froeschl{\'e}}, {Rickman}, \&
  {Valsecchi}}]{2011Icar..214..334F}
{Fouchard}, M., {Froeschl{\'e}}, C., {Rickman}, H., \& {Valsecchi}, G.~B. 2011,
  \icarus, 214, 334

\bibitem[{{Gaia Collaboration} {et~al.}(2016{\natexlab{a}}){Gaia
  Collaboration}, {Brown}, {Vallenari}, {Prusti}, {de Bruijne}, {Mignard}, \&
  et~al.}]{2016A&A...595A...2G}
{Gaia Collaboration}, {Brown}, A.~G.~A., {Vallenari}, A., {et~al.}
  2016{\natexlab{a}}, \aap, 595, A2

\bibitem[{{Gaia Collaboration} {et~al.}(2016{\natexlab{b}}){Gaia
  Collaboration}, {Prusti}, {de Bruijne}, {Brown}, {Vallenari}, {Babusiaux}, \&
  et~al.}]{2016A&A...595A...1G}
{Gaia Collaboration}, {Prusti}, T., {de Bruijne}, J.~H.~J., {et~al.}
  2016{\natexlab{b}}, \aap, 595, A1

\bibitem[{{Garc{\'{\i}}a-S{\'a}nchez}
  {et~al.}(1999){Garc{\'{\i}}a-S{\'a}nchez}, {Preston}, {Jones}, {Weissman},
  {Lestrade}, {Latham}, \& {Stefanik}}]{1999AJ....117.1042G}
{Garc{\'{\i}}a-S{\'a}nchez}, J., {Preston}, R.~A., {Jones}, D.~L., {et~al.}
  1999, \aj, 117, 1042

\bibitem[{{Garc{\'{\i}}a-S{\'a}nchez}
  {et~al.}(2001){Garc{\'{\i}}a-S{\'a}nchez}, {Weissman}, {Preston}, {Jones},
  {Lestrade}, {Latham}, {Stefanik}, \& {Paredes}}]{2001A&A...379..634G}
{Garc{\'{\i}}a-S{\'a}nchez}, J., {Weissman}, P.~R., {Preston}, R.~A., {et~al.}
  2001, \aap, 379, 634

\bibitem[{{Gontcharov}(2006)}]{2006AstL...32..759G}
{Gontcharov}, G.~A. 2006, Astronomy Letters, 32, 759

\bibitem[{{H{\o}g} {et~al.}(2000){H{\o}g}, {Fabricius}, {Makarov}, {Urban},
  {Corbin}, {Wycoff}, {Bastian}, {Schwekendiek}, \&
  {Wicenec}}]{2000A&A...355L..27H}
{H{\o}g}, E., {Fabricius}, C., {Makarov}, V.~V., {et~al.} 2000, \aap, 355, L27

\bibitem[{{Jim{\'e}nez-Torres} {et~al.}(2011){Jim{\'e}nez-Torres}, {Pichardo},
  {Lake}, \& {Throop}}]{2011MNRAS.418.1272J}
{Jim{\'e}nez-Torres}, J.~J., {Pichardo}, B., {Lake}, G., \& {Throop}, H. 2011,
  \mnras, 418, 1272

\bibitem[{{Just} {et~al.}(2015){Just}, {Fuchs}, {Jahrei{\ss}}, {Flynn},
  {Dettbarn}, \& {Rybizki}}]{2015MNRAS.451..149J}
{Just}, A., {Fuchs}, B., {Jahrei{\ss}}, H., {et~al.} 2015, \mnras, 451, 149

\bibitem[{{Kunder} {et~al.}(2017){Kunder}, {Kordopatis}, {Steinmetz},
  {Zwitter}, {McMillan}, {Casagrande}, {Enke}, {Wojno}, {Valentini},
  {Chiappini}, {Matijevi{\v c}}, {Siviero}, {de Laverny}, {Recio-Blanco},
  {Bijaoui}, {Wyse}, {Binney}, {Grebel}, {Helmi}, {Jofre}, {Antoja}, {Gilmore},
  {Siebert}, {Famaey}, {Bienaym{\'e}}, {Gibson}, {Freeman}, {Navarro},
  {Munari}, {Seabroke}, {Anguiano}, {{\v Z}erjal}, {Minchev}, {Reid},
  {Bland-Hawthorn}, {Kos}, {Sharma}, {Watson}, {Parker}, {Scholz}, {Burton},
  {Cass}, {Hartley}, {Fiegert}, {Stupar}, {Ritter}, {Hawkins}, {Gerhard},
  {Chaplin}, {Davies}, {Elsworth}, {Lund}, {Miglio}, \&
  {Mosser}}]{2017AJ....153...75K}
{Kunder}, A., {Kordopatis}, G., {Steinmetz}, M., {et~al.} 2017, \aj, 153, 75

\bibitem[{{Lindegren} {et~al.}(2016){Lindegren}, {Lammers}, {Bastian},
  {Hern{\'a}ndez}, {Klioner}, \& et~al.}]{2016A&A...595A...4L}
{Lindegren}, L., {Lammers}, U., {Bastian}, U., {et~al.} 2016, \aap, 595, A4

\bibitem[{{Malaroda} {et~al.}(2000){Malaroda}, {Levato}, {Morrell},
  {Garc{\'{\i}}a}, {Grosso}, \& {Bolzicco}}]{2000A&AS..144....1M}
{Malaroda}, S., {Levato}, H., {Morrell}, N., {et~al.} 2000, \aaps, 144, 1

\bibitem[{{Maldonado} {et~al.}(2010){Maldonado}, {Mart{\'{\i}}nez-Arn{\'a}iz},
  {Eiroa}, {Montes}, \& {Montesinos}}]{2010A&A...521A..12M}
{Maldonado}, J., {Mart{\'{\i}}nez-Arn{\'a}iz}, R.~M., {Eiroa}, C., {Montes},
  D., \& {Montesinos}, B. 2010, \aap, 521, A12

\bibitem[{{Mamajek} {et~al.}(2015){Mamajek}, {Barenfeld}, {Ivanov}, {Kniazev},
  {V{\"a}is{\"a}nen}, {Beletsky}, \& {Boffin}}]{2015ApJ...800L..17M}
{Mamajek}, E.~E., {Barenfeld}, S.~A., {Ivanov}, V.~D., {et~al.} 2015, \apjl,
  800, L17

\bibitem[{{Martell} {et~al.}(2017){Martell}, {Sharma}, {Buder}, {Duong},
  {Schlesinger}, {Simpson}, {Lind}, {Ness}, {Marshall}, {Asplund},
  {Bland-Hawthorn}, {Casey}, {De Silva}, {Freeman}, {Kos}, {Lin}, {Zucker},
  {Zwitter}, {Anguiano}, {Bacigalupo}, {Carollo}, {Casagrande}, {Da Costa},
  {Horner}, {Huber}, {Hyde}, {Kafle}, {Lewis}, {Nataf}, {Navin}, {Stello},
  {Tinney}, {Watson}, \& {Wittenmyer}}]{2017MNRAS.465.3203M}
{Martell}, S.~L., {Sharma}, S., {Buder}, S., {et~al.} 2017, \mnras, 465, 3203

\bibitem[{{Mart{\'{\i}}nez-Barbosa} {et~al.}(2017){Mart{\'{\i}}nez-Barbosa},
  {J{\'{\i}}lkov{\'a}}, {Portegies Zwart}, \& {Brown}}]{2017MNRAS.464.2290M}
{Mart{\'{\i}}nez-Barbosa}, C.~A., {J{\'{\i}}lkov{\'a}}, L., {Portegies Zwart},
  S., \& {Brown}, A.~G.~A. 2017, \mnras, 464, 2290

\bibitem[{{Matthews}(1994)}]{1994QJRAS..35....1M}
{Matthews}, R.~A.~J. 1994, \qjras, 35, 1

\bibitem[{{M{\"u}ll{\"a}ri} \& {Orlov}(1996)}]{1996EM&P...72...19M}
{M{\"u}ll{\"a}ri}, A.~A. \& {Orlov}, V.~V. 1996, Earth Moon and Planets, 72, 19

\bibitem[{{Oort}(1950)}]{1950BAN....11...91O}
{Oort}, J.~H. 1950, \bain, 11, 91

\bibitem[{{Pasquini} {et~al.}(2011){Pasquini}, {Melo}, {Chavero}, {Dravins},
  {Ludwig}, {Bonifacio}, \& {de La Reza}}]{2011A&A...526A.127P}
{Pasquini}, L., {Melo}, C., {Chavero}, C., {et~al.} 2011, \aap, 526, A127

\bibitem[{{Pourbaix}(2011)}]{2011AIPC.1346..122P}
{Pourbaix}, D. 2011, in American Institute of Physics Conference Series, Vol.
  1346, American Institute of Physics Conference Series, ed. J.~A. {Docobo},
  V.~S. {Tamazian}, \& Y.~Y. {Balega}, 122--133

\bibitem[{{Rickman}(1976)}]{1976BAICz..27...92R}
{Rickman}, H. 1976, Bulletin of the Astronomical Institutes of Czechoslovakia,
  27, 92

\bibitem[{{Rickman} {et~al.}(2012){Rickman}, {Fouchard}, {Froeschl{\'e}}, \&
  {Valsecchi}}]{2012P&SS...73..124R}
{Rickman}, H., {Fouchard}, M., {Froeschl{\'e}}, C., \& {Valsecchi}, G.~B. 2012,
  \planss, 73, 124

\bibitem[{{Scholl} {et~al.}(1982){Scholl}, {Cazenave}, \&
  {Brahic}}]{1982A&A...112..157S}
{Scholl}, H., {Cazenave}, A., \& {Brahic}, A. 1982, \aap, 112, 157

\bibitem[{{SDSS Collaboration} {et~al.}(2016){SDSS Collaboration}, {Albareti},
  {Allende Prieto}, {Almeida}, {Anders}, {Anderson}, {Andrews},
  {Aragon-Salamanca}, {Argudo-Fernandez}, {Armengaud}, \&
  et~al.}]{2016arXiv160802013S}
{SDSS Collaboration}, {Albareti}, F.~D., {Allende Prieto}, C., {et~al.} 2016,
  ArXiv e-prints

\bibitem[{{van Leeuwen}(2007)}]{2007ASSL..350.....V}
{van Leeuwen}, F., ed. 2007, Astrophysics and Space Science Library, Vol. 350,
  {Hipparcos, the New Reduction of the Raw Data}

\bibitem[{{Weissman}(1996)}]{1996EM&P...72...25W}
{Weissman}, P.~R. 1996, Earth Moon and Planets, 72, 25

\bibitem[{Öpik(1932)}]{10.2307/20022899}
Öpik, E. 1932, Proceedings of the American Academy of Arts and Sciences, 67,
  169

\end{thebibliography}
